\documentclass[useAMS,usenatbib]{mn2e}
\usepackage{graphicx,amsmath,amssymb}
\usepackage{natbib}
\newcommand{\comment}[1]{}

\def\simgt{\lower.5ex\hbox{$\; \buildrel > \over \sim \;$}}
\def\simlt{\lower.5ex\hbox{$\; \buildrel < \over \sim \;$}}

\title[Stardust from AGB and Super--AGB stars]{The transition from carbon dust to silicates 
production in low-metallicity AGB and SAGB stars}
\author[Ventura et al.]{P. Ventura,$^1$ M. Di Criscienzo,$^1$ R. Schneider,$^1$ 
R. Carini,$^{1,2}$ R. Valiante,$^{1,3}$ 
\newauthor
F. D'Antona,$^1$ S. Gallerani,$^1$ R. Maiolino,$^1$ A. Tornamb\'e$^1$  \\
$^1$INAF -- Osservatorio Astronomico di Roma, Via Frascati 33, 00040, Monte Porzio Catone (RM), Italy \\
$^{2}$Dipartimento di Fisica, Universit\`a di Roma ``La Sapienza'', P.le Aldo Moro 5, 00143, 
Roma, Italy \\
$^{3}$INAF -- Osservatorio Astrofisico di Arcetri, Largo Enrico Fermi 5, 50125 Firenze, Italy}
\begin{document}

\date{Accepted, Received; in original form }

\pagerange{\pageref{firstpage}--\pageref{lastpage}} \pubyear{2011}

\maketitle

\label{firstpage}

\begin{abstract}

We compute the mass and composition of dust produced by stars
with masses in the range $1 M_{\odot} \le M \le 8 M_{\odot}$ and with a
metallicity of $Z = 0.001$ during their AGB and Super AGB phases.
Stellar evolution is followed from the
pre-main sequence phase using the code ATON which provides, at each timestep,
the thermodynamics and the chemical stucture of the wind. 
We use a simple model to describe the growth of the dust grains under the hypothesis of a 
time--independent, spherically symmetric stellar wind. Although part of the modelling
which describes the stellar outflow is not completely realistic, this approach allows a
straight comparison with results based on similar assumptions present in the literature, and
thus can be used as an indication of the uncertainties affecting the theoretical investigations 
focused on the dust formation process in the surroundings of AGB stars.

We find that the total mass of dust injected by AGB stars 
in the interstellar medium does not increase monotonically with stellar mass and 
ranges between a minimum of $10^{-6} M_{\odot}$ for the $1.5 M_{\odot}$ stellar model,
up to $2 \times 10^{-4} M_{\odot}$, for the $6 M_{\odot}$ case. 
Dust composition depends on the stellar mass:   
low--mass stars ($M < 3M_{\odot}$) produce carbon--rich dust, 
whereas more massive stars, experiencing Hot Bottom Burning, never reach 
the carbon--star stage, and produce silicates and iron. This is in partial
disagreement with previous investigations in the literature, which are
based on synthetic AGB models and predict 
that, when the initial metallicity is $Z=0.001$, C--rich dust is 
formed at all stellar masses. The differences are due to 
the different modelling of turbulent convection in the super--adiabaticity 
regime. Also in this case, like for other physical features of the AGB, the treatment of
super--adiabatic convection shows up as the most relevant issue affecting the dust--formation 
process.

We also investigate Super AGB stars with masses
$6.5 M_{\odot} \le M \le 8 M_{\odot}$ that evolve over a ONe core.
Due to a favourable combination of mass loss and Hot Bottom Burning,
these stars are predicted to be the most efficient silicate--dust
producers, releasing $[2 - 7]\times 10^{-4} M_{\odot}$ masses of
dust. 

We discuss the robustness of these predictions and their relevance
for the nature and evolution of dust at early cosmic times.
\end{abstract}


\section{Introduction}
Winds from stars on the asymptotic giant branch (AGB) and super-asymptotic giant branch (SAGB) 
provide an important component of mass return into the interstellar medium (ISM)
and account for a significant fraction of interstellar dust in present-day mature galaxies 
(Zhukovska, Gail \& Trieloff 2008; Sloan et al. 2009). Recent chemical evolutionary models, which
attempt to constrain the origin of the large dust masses inferred from mm and submm data
of $z > 6$ QSOs, have shown that AGB stars give important contributions to the dust
content even at these early cosmic epochs (Valiante et al. 2009, 2011). 

Observations with the {\it Spitzer} telescope have probed the formation of dust
in AGB stars of galaxies in the Local Group (Sloan et al. 2008, 2010a and references therein)
and in globular clusters (Sloan et al. 2010b; McDonald et al. 2011). 
In spite of continuous progress, detailed understanding of the outflow dynamics and
grain formation is far from being complete (for a comprehensive review 
see H$\ddot{o}$fner 2009): dust grain properties depend on dynamical aspects 
of the stellar wind, and on the surface chemical composition of the star. Shock
waves caused by stellar pulsations lift the gas above the stellar surface intermittently
creating dense cool layers where grain condensation can occurr. Provided that their
cross sections are large enough, the newly formed dust grains can be accelerated by 
radiation pressure and collisionally couple to the gas, dragging it along.
Time--independent wind models, though limited in their capability
of reproducing important properties of the dynamical atmosphere with shocks, can reproduce
certain observed features, such as the observed spectral energy distribution of
C--rich stars with high mass loss rates.

Compared to the problem of computing reliable mass loss rates, the problem of calculating
dust yields has the additional complication of estimating the fraction of elements
condensing into various types of dust species, which, in turn, depends on the changes
in the surface chemistry of the star determined by the interplay between nucleosynthesis
and convection.

The most extensive study so far has been made by Ferrarotti \& Gail (2006). Using synthetic 
stellar evolution models, they compute the 
non-equilibrium dust formation and estimate dust yields for M--, S--, and C--type AGBs. 
Synthetic AGB models are useful tools, due to the relatively short computation time required and
the simplicity with which they can be incorporated in stellar population synthesis studies.
On the other hand, they have no predictive power on some key physical processes that occurr
during AGB evolution, such as the Hot Bottom Burning experienced at the bottom of the convective 
envelope. A proper description of these processes and of their impact on dust formation 
demands the full integration of the stellar structure.

To progress in this direction, the aim of the present investigation is to 
estimate the mass and composition of dust formed around AGBs using models that
follow stellar evolution from the pre--main sequence phase. This enables the
computation of the time evolution of the surface chemistry and of the mass loss rate in a 
self--consistent way, providing an important benchmark for studies based on synthetic modelling. 
Using the proper stellar evolutionary sequences, we also apply the dust formation model to 
winds of SAGB stars, whose dust yields have not been studied so far.

The paper is organized as follows. The description of the codes used to calculate the AGB and
SAGB evolution, the wind structure and dust formation are described in Sect.2; the main
physical processes which affect the surface chemistry of these stars are discussed
in Sect.3. In Sect.4 we present the resulting dust mass and composition for stars with
masses in the range $1 M_{\odot} \le M \le 8 M_{\odot}$ and with initial 
metallicity of $Z = 0.001$; we assess the robustness of these results and discuss the effects
of uncertain physical processes in Sect.5. Finally, in Sect. 6 we compare our predictions with
alternative dust yields present in the literature and in Sect. 7 we summarize the main conclusions.

\section{Physical and numerical inputs}
\label{sec:refmodel}
To address the issue of dust production by AGB stars, we developed a specific
tool to calculate the structure of the wind around the central object as
well as the mass and composition of the newly formed dust. These properties 
depend on the luminosity, effective temperature and mass loss rate experienced 
by the star, and on the surface chemical composition.  
In the following, we provide a description of the numerical code used to integrate
the stellar evolution, the ATON code, and of the code used to estimate the stellar wind 
structure and dust formation.

\subsection{The ATON evolution code}
The evolution of the stars, from their pre Main Sequence phase, until the almost
complete ejection of their external mantle, was computed by means of the code
ATON \citep{ventura98} which integrates spherically symmetric structures in 
hydrostatic equilibrium. An exaustive description of the code is beyond the 
scope of the present study and we refer the interested reader to the original
paper. Here we briefly recall those features which affect the thermodynamics
and chemical composition of the wind where dust nucleation occurs.

\subsubsection{The convection model}
The temperature gradient within regions unstable to convection is determined via the
Full Spectrum of Turbulence (\citet{cm91}, hereinafter FST) model for turbulent convection; 
the code allows the alternative Mixing Length Theory (MLT) treatment \citep{vitense}.
Mixing of chemicals within nuclearly active regions is handled by means of a
diffusive approach: for each of the species included in the nuclear network we solve
the diffusive equation following the scheme by \citet{cloutman}, which couples 
self--consistently nuclear burning and mixing of chemicals. 
The extra--mixing (overshooting) from the convective borders is simulated by an exponential decay of
velocities from the neutrality point, where buoyancy vanishes; the scale of this
phenomenon is assumed to be $l=\zeta H_p$, where $H_p$ is the pressure scale height.
In agreement with the calibration based on the fit of the observed main sequences of open 
clusters given in  \citet{ventura98}, we assume $\zeta=0.02$ to simulate overshoot from 
the border of the
convective cores during the two major phases of core hydrogen and helium burning.
To limit the number of free parameters, we assumed no extra-mixing from the base of
the convective envelope during the giant evolution, and during the AGB phase. This
choice renders the extent of the Third Dredge Up found during the Thermal Pulses phase
a conservative estimate.

\subsubsection{The mass loss rate}
The mass loss rate during the AGB evolution was modelled according to \citet{blocker95}.
Based on hydrodynamic computations by \citet{bowen}, the strong increase in the mass loss
rate during the AGB evolution is modelled by multiplying the Reimers' rate by a given 
power of the luminosity; the resulting expression is,

\begin{equation}
\dot M=4.83 \times 10^{-22} \eta_R M^{-3.1}L^{3.7}R
\label{blocker}
\end{equation}
\noindent
where $\eta_R$ is a free parameter and $M$, $R$, and $L$ are the stellar
mass, radius and luminosity, expressed in solar units. Using as a calibration the luminosity
function of lithium--rich stars in the Magellanic Clouds, we set $\eta_R=0.02$ \citep{ventura00}.

The Bl\"ocker treatment has been used and tested so far to describe mass loss from 
high--luminosity, M stars; here we extend this modelling to the C--rich regime, being
aware of the additional uncertainties due to this choice on the results obtained. 
\citet{mattsson07} present an analysis of the uncertainties in the description of 
the mass loss during the AGB phase of carbon stars (see in particular their Fig.~4).

Some sequences based on the \citet{VW93} treatment (hereinafter VW93) were also calculated 
for comparison: in this case mass loss increases exponentially 
with the pulsation period, P, until the star enters a super--wind phase, when P 
exceeds 500d. The super--wind rate is $\dot M=5\times 10^{-5}$M$_{\odot}$/yr.

The results provided by the two descriptions are in reasonable agreement for low--mass AGBs, 
and for the early AGB phases of the more massive objects; the main differences are found
in the description of the advanced phases of the evolution of the stars with mass close to
the limit for carbon ignition, where the \citet{blocker95} treatment predicts very large
mass loss rates, that favour an earlier consumption of the whole stellar envelope.

\subsubsection{Nuclear network and opacities}
The nuclear network includes 30 elements, with 64 reactions. All the most
relevant p--capture and $\alpha$--capture processes are taken into account. The nuclear
cross-sections were taken from the NACRE compilation \citep{angulo}, with the exception of 
the reactions $^{14}$N(p,$\gamma)^{15}$O \citep{formicola},
$^{12}$C($\alpha,\gamma)^{16}$O \citep{kunz}, 
$3\alpha \longrightarrow ^{12}$C+$\gamma$ \citep{fynbo}, 
$^{22}$Ne(p,$\gamma)^{23}$Na \citep{hale02}, $^{23}$Na(p,$\gamma)^{24}$Mg \citep{hale04},
$^{23}$Na(p,$\alpha)^{20}$Ne \citep{hale04}.

The opacities in the low-temperature regime are calculated by means of the AESOPUS tool,
developed by \citet{marigo09}. These tables are suitably constructed to follow the changes 
in the chemical composition of the envelope when the C/O ratio approaches, or exceeds, unity.
This is deeply different from the classic approach, traditionally used to build AGB
models, that neglects possible variations in the surface chemistry during the evolution in
the opacity calculations. To understand how the results depend on this choice, we decided to
calculate additional AGB models, where the above effect is neglected. It goes without saying
that this is expeted to play a role only in the models that eventually reach the C--star 
stage, i.e. for M$<4$M$_{\odot}$ in the present investigation.

For the present study, we restrict our discussion to stellar models with absolute metallicity
Z=0.001, and helium fraction Y=0.24. The mixture is $\alpha$--enhanced with $[\alpha$/Fe]=+0.4, 
the solar reference mixture being \citet{gs98}. Other chemical compositions will be
investigated in future explorations.

\subsection{Stellar wind and dust formation}
The thermodynamic and chemical structure of the wind that forms around AGBs is 
calculated on the basis of the inputs provided at each time step by the integration of the
structure of the central star, which are the natural outcomes of the ATON code. In
particular, we use luminosity ($L$), effective temperature ($T_{\rm eff}$), mass ($M$), 
mass loss rate ($\dot M$), and surface chemistry defined by the mass fractions ($X_i$)
of the various elements included in the nuclear network.

\subsubsection{The structure of the wind}
The structure of the wind is determined following the schematization by \citet{fg06}.

The outflow is assumed to be stationary and spherically symmetric. Neglecting the 
hydrostatic pressure forces, we find from the equation for momentum conservation that,
\begin{equation}
v{dv\over dr}=-{GM\over r^2}(1-\Gamma).
\label{eqv}
\end{equation}
\noindent
The quantity $\Gamma$ gives the ratio between the radiative pressure on the dust and
the gravitational pull and is given by,

\begin{equation}
\Gamma={kL\over 4\pi cGM},
\label{eqgamma}
\end{equation}
\noindent
where $k$ is the flux--averaged extinction coefficient of the gas--dust
mixture and can be expressed as,

\begin{equation}
k=k_{\rm gas}+\sum_i f_ik_i,
\label{eqk}
\end{equation}
\noindent
with $k_{\rm gas}=10^{-8}\rho^{2/3}T^3$ \citep{bell94}. The sum in eq.~(\ref{eqk}) is 
extended to all the dust species considered: the $f_i$ terms give the degrees of condensation 
of the key--elements for each dust species and $k_i$ represent their corresponding extinction 
coefficients. 

To completely define the thermal structure of the wind, we need to specify the density and
temperature radial profiles. Mass conservation leads to,

\begin{equation}
\dot{M}=4\pi r^2 \rho v,
\label{eqrho}
\end{equation}
\noindent
and the temperature structure is determined by adopting the approximation by 
\citet{lucy71,lucy76} that holds for spherically symmetric, grey winds:

\begin{equation}
T^4={1\over 2}T_{\rm eff}^4 \left[ 1-\sqrt{1-{R^2\over r^2}}+{3\over 2}\tau \right].
\label{eqt}
\end{equation}
\noindent
The optical depth $\tau$ in eq.~(\ref{eqt}) is found from the differential relation,

\begin{equation}
{d\tau\over dr}=-\rho k{R^2\over r^2}
\label{eqtau}
\end{equation}
\noindent
with the limiting condition that $\tau \longrightarrow 0$ for $r \longrightarrow \infty$.

The wind model, determined via integration of eqs.~(\ref{eqv})--(\ref{eqtau}), is calculated
starting from the innermost radius where the first dust species become stable, commonly
located 3--4 stellar radii away from the centre of the star. This is
expected to coincide with the region where $\Gamma$ increases until it exceeds unity, and
the wind is accelerated to supersonic velocities. In analogy with \citet{fg06}, we keep the
velocity of the gas constant from the surface of the star to the region where dust forms.
The above equations are integrated out to a radial distance of $10^4 R$ ($R$ is the stellar 
radius), far beyond the point where all the relevant quantities reach their asymptotic behaviour.

Before turning to the dust formation process, we stress here that a more physically sound
description of the wind properties is needed before the full reliability of the results 
from this kind of investigations can be confirmed. In particular, the assumption of a 
time--independent flow neglects any effect of atmospheric shocks waves, which play a crucial
role for the wind mechanism. The assumption of a constant velocity does not account for the
consequences of these shocks, which could push the gas beyond the condensation point.

\subsubsection{Dust formation}
Dust formation generally starts with the condensation of small seed nuclei which 
then grow by accreting material to macroscopic grains. Since the nature of these seeds
is still unclear, we do not consider the nucleation phase, but only the
growth process of grains. As initial conditions, in agreement with \citet{fg06}, we 
assume that the seed grains have an initial size $a_0=1$nm, and that their initial
density is $n_d= 3 \times 10^{-4}n_H$, where $n_H$ is the hydrogen number density in the 
wind. We could verify that the results are independent of both assumptions.

The composition of dust in the winds during the AGB phase depends on the surface chemical
composition. For M stars, with surface C/O $<$ 1, the dominant dust species are expected to
be olivine, pyroxene, quartz, and solid iron. For C stars with surface C/O $>1$, the dust
mixture is assumed to be dominated by solid carbon, SiC and solid iron. In Table 1 we 
summarize the formation reactions of these dust species. When computing the reaction rates for 
olivine and pyroxene, we follow the variation of the magnesium percentage with respect to the pure 
iron component.

\begin{table*}
\begin{center}
\caption{Dust species considered in the present analysis, their formation reaction and
adopted sticking coefficient (see text).} 
\begin{tabular}{l|l|l|l}
\hline
\hline 
Grain Species & Formation Reaction & Sticking Coefficient \\
\hline
Olivine & 2$x$Mg +2(1-$x$)Fe+SiO+3H$_2$O $\rightarrow$ Mg$_{2x}$Fe$_{2(1-x)}$SiO$_4$ + 3H$_2$ & 0.1 \\ 
Pyroxene & $x$Mg +(1-$x$)Fe+SiO+2H$_2$O  $\rightarrow$ Mg$_{x}$Fe$_{(1-x)}$SiO$_3$ + 2H$_2$ & 0.1 \\
Quartz & SiO + H$_2$O $\rightarrow$ SiO$_2(s)$ +H$_2$ & 0.1 \\  
Silicon Carbide & 2Si + C$_2$H$_2$ $\rightarrow$ 2 SiC + H$_2$ & 1 \\ 
Carbon & C $\rightarrow$ C$(s)$ & 1 \\
Iron & Fe $\rightarrow$ Fe$(s)$ & 1 \\
\hline 
\hline 
\end{tabular}
\end{center}
\label{tabrates}
\end{table*}

The temporal variation of the dust grain size for the $i$--th dust species, $a_i$, is
computed as a competition between the growth rate $J_i^{\rm gr}$ and the destruction 
rate $J_i^{\rm dec}$, i.e.
\begin{equation}
da_i/dt=V_{0,i}(J_i^{\rm gr}-J_i^{\rm dec}),  
\label{eq:ai}
\end{equation}
\noindent 
where $V_{0,i}$ is the volume of the nominal molecule in the solid. 
For each dust species, the growth rate is determined by the addition
rate of a key element, which is generally the least abundant species
involved in the chemical reaction. For silicates and SiC these are 
supposed to be the SiO and C$_2$H$_2$ molecules, respectively, whereas 
for solid iron and carbon it is the corresponding element in the gas phase. 
The growth rates per unit time and surface area are proportional to the
thermal velocity, $v_{th}$, and density, $n$, of the key species in the gas phase,
\begin{equation}
J_{i}^{\rm gr} =  \alpha_{i} n_{i} v_{th,i},
\end{equation}
\noindent
where the index $i$ runs over the different key species and $\alpha$ is the sticking coefficient. 

Vaporization of dust grains by thermal decomposition is estimated from the vapour pressure
of the key species, $p_{v,i}$, over the solid state, and it is also proportional to the thermal
velocity, reading
\begin{equation}
J_{i}^{\rm dec} = \alpha_{i} v_{th,i} \frac{p_{v,i}}{kT}.
\end{equation}
\noindent
For carbon, based on the discussion in \citet{fg06}, we assume that condensation
begins at a temperature of 1100~K and that the destruction term is negligible.

The vapour pressures $p_{v,i}$ in Eq.~10 are found via the law of mass action: it 
allows to express the individual equilibrium gas pressures of the molecules involved  
on the basis of the total variation of enthalpy associated to the dust formation
reaction. An exhaustive description of the methodology followed to estimate the
individual vapour pressures can be found, e.g., in \citet{fg01}.

The formation of the four dust species in M--stars (C/O $<1$) is therefore characterized by 
the following four equations,
\begin{equation}
{da_{\rm ol}\over dt}=V_{0,\rm ol}(J_{\rm ol}^{\rm gr}-J_{\rm ol}^{\rm dec}),
\label{eqol}
\end{equation}

\begin{equation}
{da_{\rm py}\over dt}=V_{0,\rm py}(J_{\rm py}^{\rm gr}-J_{\rm py}^{\rm dec}),
\label{eqpy}
\end{equation}

\begin{equation}
{da_{\rm qu}\over dt}=V_{0,\rm qu}(J_{\rm qu}^{\rm gr}-J_{\rm qu}^{\rm dec}),
\label{eqqu}
\end{equation}

\begin{equation}
{da_{\rm ir}\over dt}=V_{0,\rm ir}(J_{\rm ir}^{\rm gr}-J_{\rm ir}^{\rm dec}).
\label{eqir}
\end{equation}
\noindent
These equations are completed by two further relations, giving the percentages of
magnesium in the olivine and pyroxene dust particles:

\begin{equation}
{dx_{\rm ol}\over dt}={3V_{0,\rm ol}\over a_{ol}}\left[ 
(x_g-x_{\rm ol})J_{\rm ol}^{\rm gr}+{1\over 2}(J_+^{\rm ex}-J_-^{\rm ex}) \right]
\label{eqxol}
\end{equation}

\begin{equation}
{dx_{\rm py}\over dt}={3V_{0,\rm py}\over a_{py}}\left[ 
(x_g-x_{\rm py})J_{\rm py}^{\rm gr}+{1\over 2}(J_+^{\rm ex}-J_-^{\rm ex}) \right]
\label{eqxpy}
\end{equation}
\noindent
where $x_g$ is defined as the relative gas abundance of magnesium with respect to the Mg+Fe sum.

The quantity $(J_+^{\rm ex}-J_-^{\rm ex})$ gives the difference between the exchange
rate of iron by magnesium per unit surface area during collisions of Mg with the grain
surface and the rate of the reverse reaction, and is given by

\begin{equation}
(J_+^{\rm ex}-J_-^{\rm ex})=v_{th,Mg}\alpha^{ex}(n_{Mg}-n_{ir}K_p) 
\label{exch}
\end{equation}

where $K_p$ is calculated on the basis of the free enthalpies of formation of the
pure products with only magnesium or iron, and $\alpha^{ex}$ is the exchange coefficient,
describing the probability of exchange of magnesium and iron in the scattering process.

Dust formation in C--stars is described by eq.~(\ref{eqir}),
plus two additional equations for carbon and SiC grains,

\begin{equation}
{da_{\rm SiC}\over dt}=V_{0,\rm SiC}(J_{\rm SiC}^{\rm gr}-J_{\rm SiC}^{\rm dec}),
\label{eqsic}
\end{equation}

\begin{equation}
{da_{\rm C}\over dt}=V_{0,\rm C}J_{\rm C}^{\rm gr}.
\label{eqc}
\end{equation}
\noindent
The growth rate coefficients $J_{\rm ol}^{\rm gr}$, $J_{\rm py}^{\rm gr}$,
$J_{\rm qu}^{\rm gr}$, $J_{\rm ir}^{\rm gr}$, $J_{\rm SiC}^{\rm gr}$,
$J_{\rm C}^{\rm gr}$, and the exchange coefficients $J_+^{\rm ex}$, $J_-^{\rm ex}$
for olyvine and pyroxene are calculated following \citet{gs99}, \citet{fg01}, \citet{fg02}
and \citet{fg06}. The sticking coefficients for iron, carbon and SiC formation are
assumed to be $\alpha=1$, whereas for olivine we take $\alpha=0.1$ (see \citet{fg02}, Table 3).
For pyroxene and quartz the sticking coefficients are less well determined from laboratory
experiments; in analogy with \citet{fg01}, we assume $\alpha_{\rm py}=\alpha_{\rm qu}=0.1$ (see the
last column of Table 1). 
Finally, the exchange coefficient entering the 
equation for the determination of the fraction of magnesium within olivine and pyroxene 
is taken to be $\alpha^{\rm ex}=0.06$.

The free enthalpy of formation of the molecules involved in the reactions for the
condensation of the various species of dust were taken from \citet{sharp99}, with the
only exceptions of FeSiO$_3$, graphite, and solid SiC, that were kindly provided by
Prof. Gail (private communication). 

The extinction coefficients for iron and quartz were calculated according to the
analytic expansion given in \citet{gs99}, whereas for olivine and pyroxene we used the
opacities derived by \citet{ossenkopf92}. For what concerns the C--rich environment, we use 
for SiC eq.~(34) in \citet{fg02}, whereas the carbon extinction coefficient was kindly 
provided by Prof. Gail (private communication). In the cases where analytic expansions
associated to solar abundaces were used, the coefficients were properly scaled, to account
for the difference in the number density of the key--species involved in the present
investigation, compared to the solar values.

\section{The change in the surface chemistry of AGB and SAGB stars}
After the core He--burning phase is finished, stars with mass 1M$_{\odot}\leq M \leq M_{\rm up}$ 
evolve through the AGB phase (see e.g. \citet{herwig00}): they experience a series of thermal 
pulses (TP), triggered
by the periodic ignition of a He--rich layer below the CNO burning shell, and end--up
as CO White Dwarfs. More massive objects, with M$_{\rm up}<$M$<$M$_{\rm ccSN}$, achieve
carbon burning in an off-center, partially degenerate, region; after experiencing some TPs,
they evolve as White Dwarfs made up of oxygen and neon. Their evolution is commonly
referred to as the Super Asymptotic Giant Branch (SAGB) phase.

The exact values of M$_{\rm up}$ and M$_{\rm ccSN}$ are determined by the core 
overshooting during the two major phases of nuclear burning; in the present investigation,
based on the assumption of a moderate extra--mixing from the outer border of the convective
core, we find $M_{\rm up}=6M_{\odot}$ and $M_{\rm ccSN}=8M_{\odot}$.

The mass and composition of dust produced by AGBs and SAGBs is determined both by the physical 
behaviour of the star and by the surface chemistry. The amount of dust produced will depend on 
the number density of the key species, which demands a detailed understanding of the 
evolution of the surface mass fraction of silicon, oxygen, carbon, magnesium and iron.

\begin{figure*}
\begin{minipage}{0.33\textwidth}
\resizebox{1.\hsize}{!}{\includegraphics{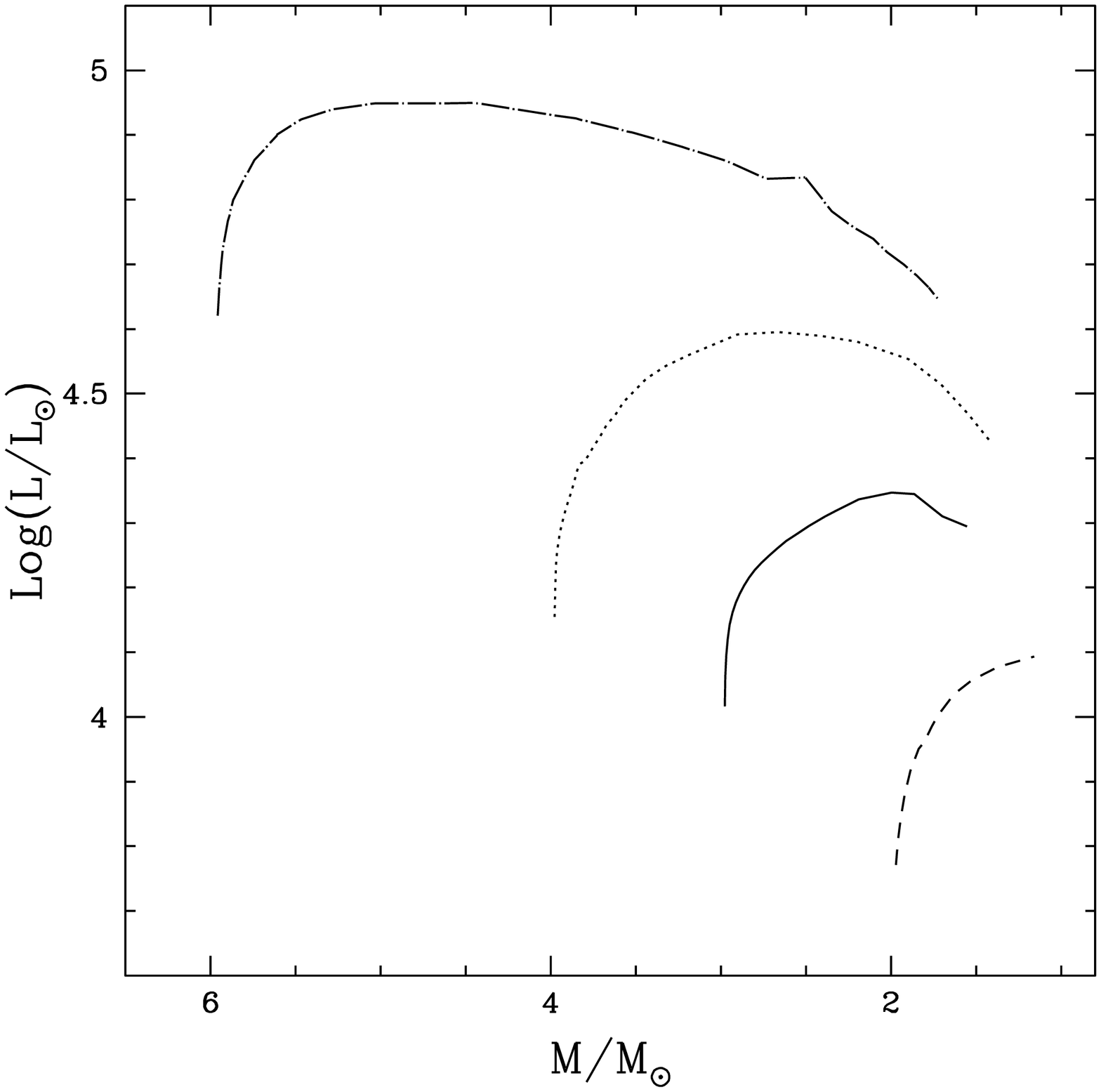}}
\end{minipage}
\begin{minipage}{0.33\textwidth}
\resizebox{1.\hsize}{!}{\includegraphics{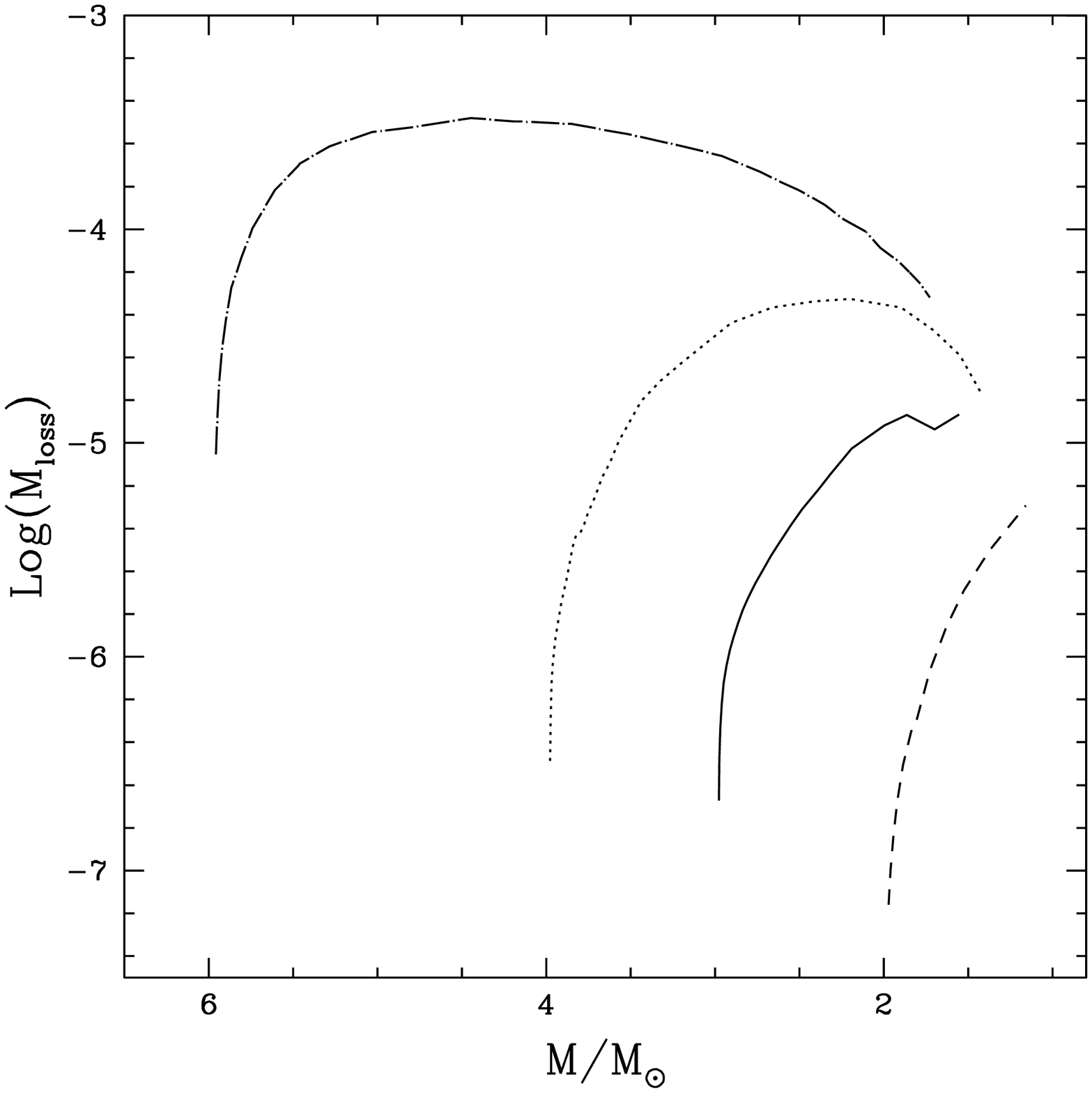}}
\end{minipage}
\begin{minipage}{0.33\textwidth}
\resizebox{1.\hsize}{!}{\includegraphics{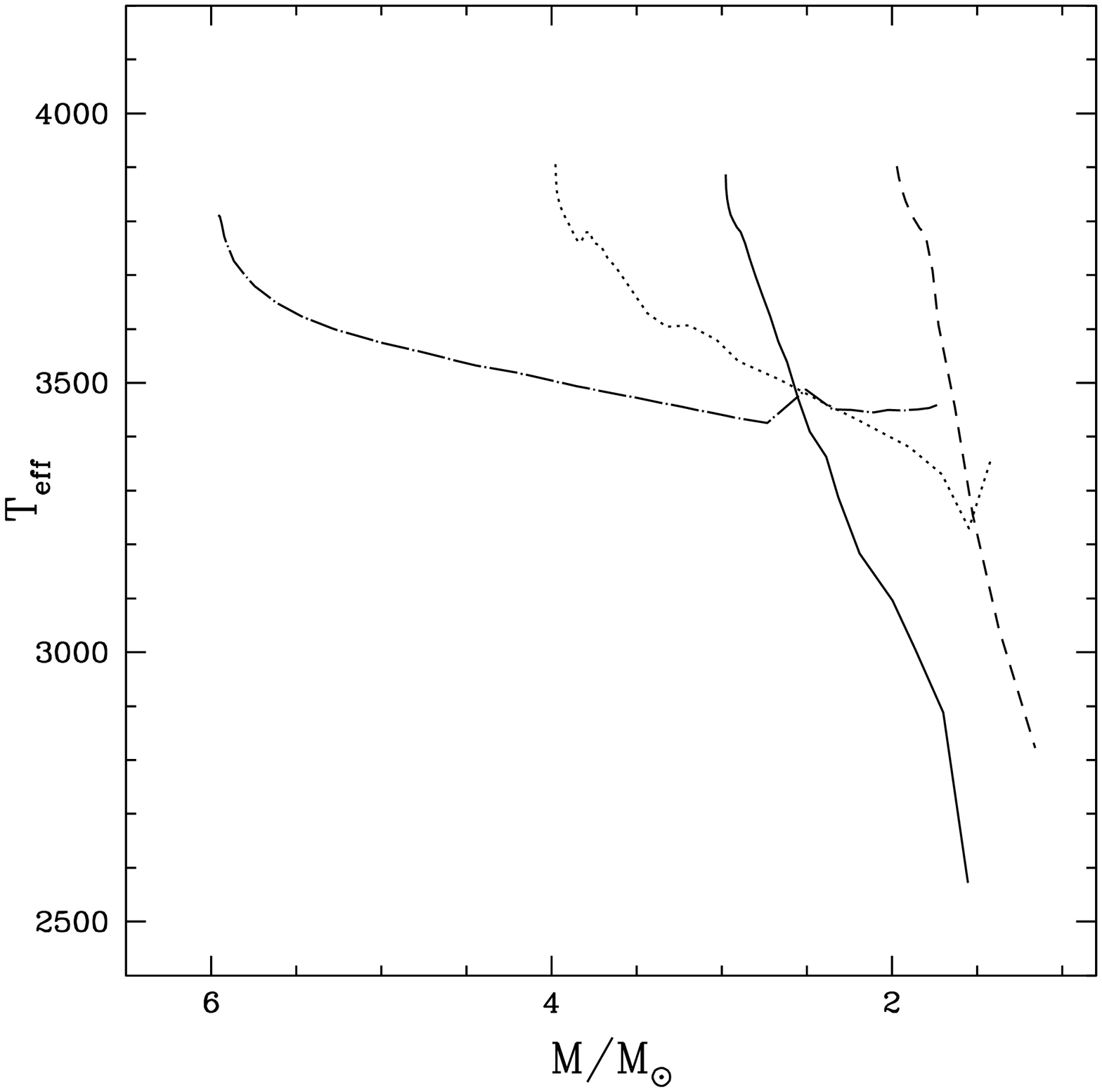}}
\end{minipage}
\caption{The evolution of the luminosity (Left), mass loss rate (Middle, expressed in
M$_{\odot}$/yr) and
effective temperature (Right) with the total mass of the star for 4 models 
with initial mass 6M$_{\odot}$ (dashed--dotted), 4M$_{\odot}$ (dotted),
3M$_{\odot}$ (solid), and 2M$_{\odot}$ (dashed). The low surface temperatures
reached by the lower mass models are favoured by the general expansion of the
whole convective envelope, that follows the achievement of the C--star stage.
}
\label{figfis}
\end{figure*}

The surface chemistry of AGBs is affected by two fundamental physical processes, 
the Third Dredge Up and the Hot Bottom Burning (e.g. \citet{iben83}). In what follows, we 
briefly describe the effects of these two mechanisms, and their associated uncertainties. 

\subsection{Third Dredge Up} 
The Third Dredge Up (hereafter TDU) 
is associated to the inwards penetration of the surface convective zone 
following each thermal pulse (TP), when the bottom of the external mantle crosses the 
entropy barrier due to the H-He discontinuity; the envelope enters a region where a 
pulse--driven convective zone, formed as a consequence of the 3$\alpha$ reactions ignition, 
had spread the products of helium nucleosynthesis; the outcome of such a mechanism is 
surface carbon enrichment, and the possible achievement of the C--star regime. The TDU 
modelling is extremely sensitive to the numerics adopted \citep{straniero97}: a narrow 
temporal and spatial zoning, with time--steps of the order of hours, and $\sim 5000$ 
mesh--points, are required to allow a self--consistent description of the TDU phenomenology.
A deeper TDU can be obtained either by assuming
some overshoot from the bottom of the surface convective zone \citep{herwig00}, or from
the borders of the convective shell that forms when the thermal pulse starts \citep{herwig04}.

\begin{figure*}
\begin{minipage}{0.33\textwidth}
\resizebox{1.\hsize}{!}{\includegraphics{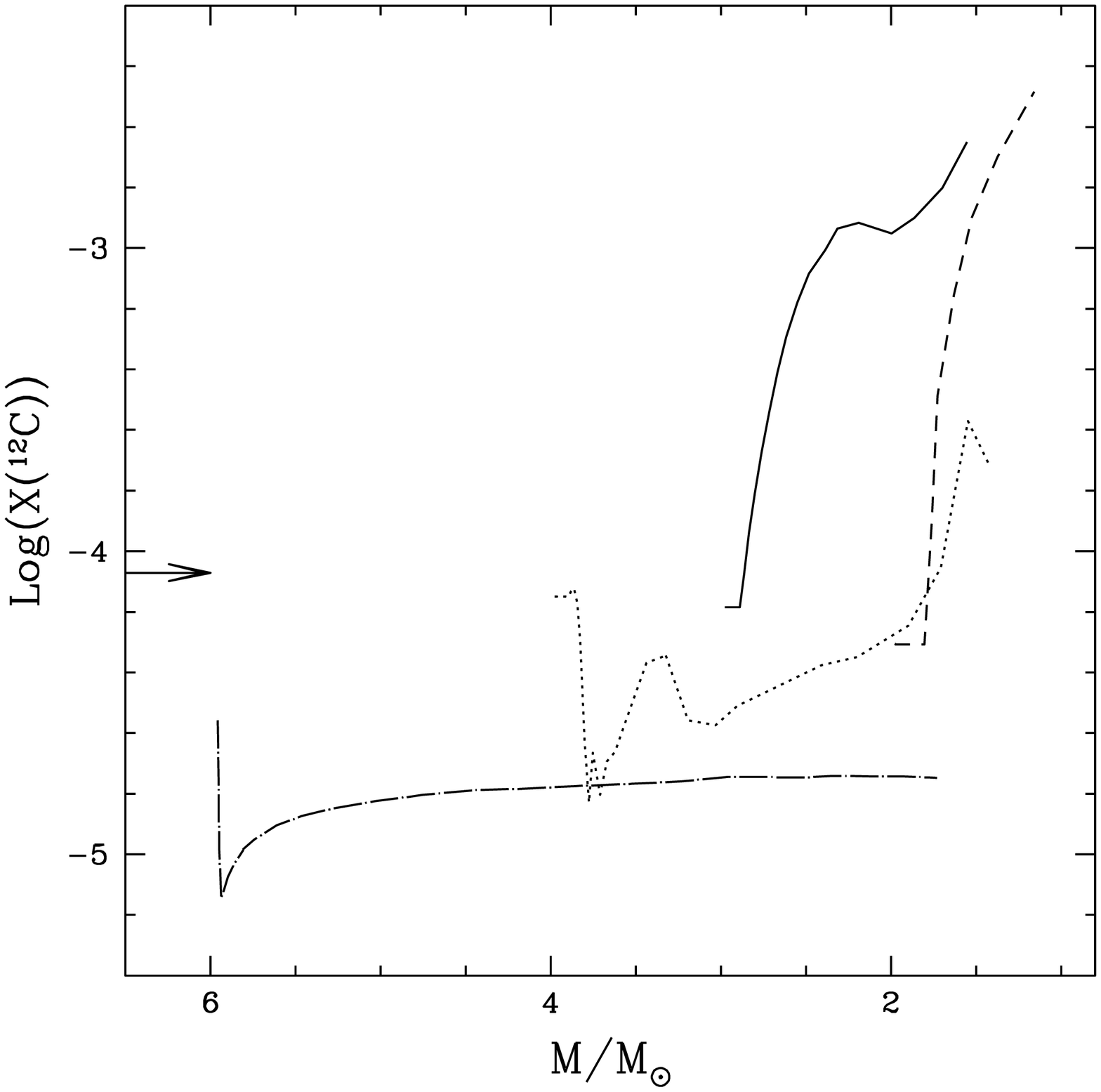}}
\end{minipage}
\begin{minipage}{0.33\textwidth}
\resizebox{1.\hsize}{!}{\includegraphics{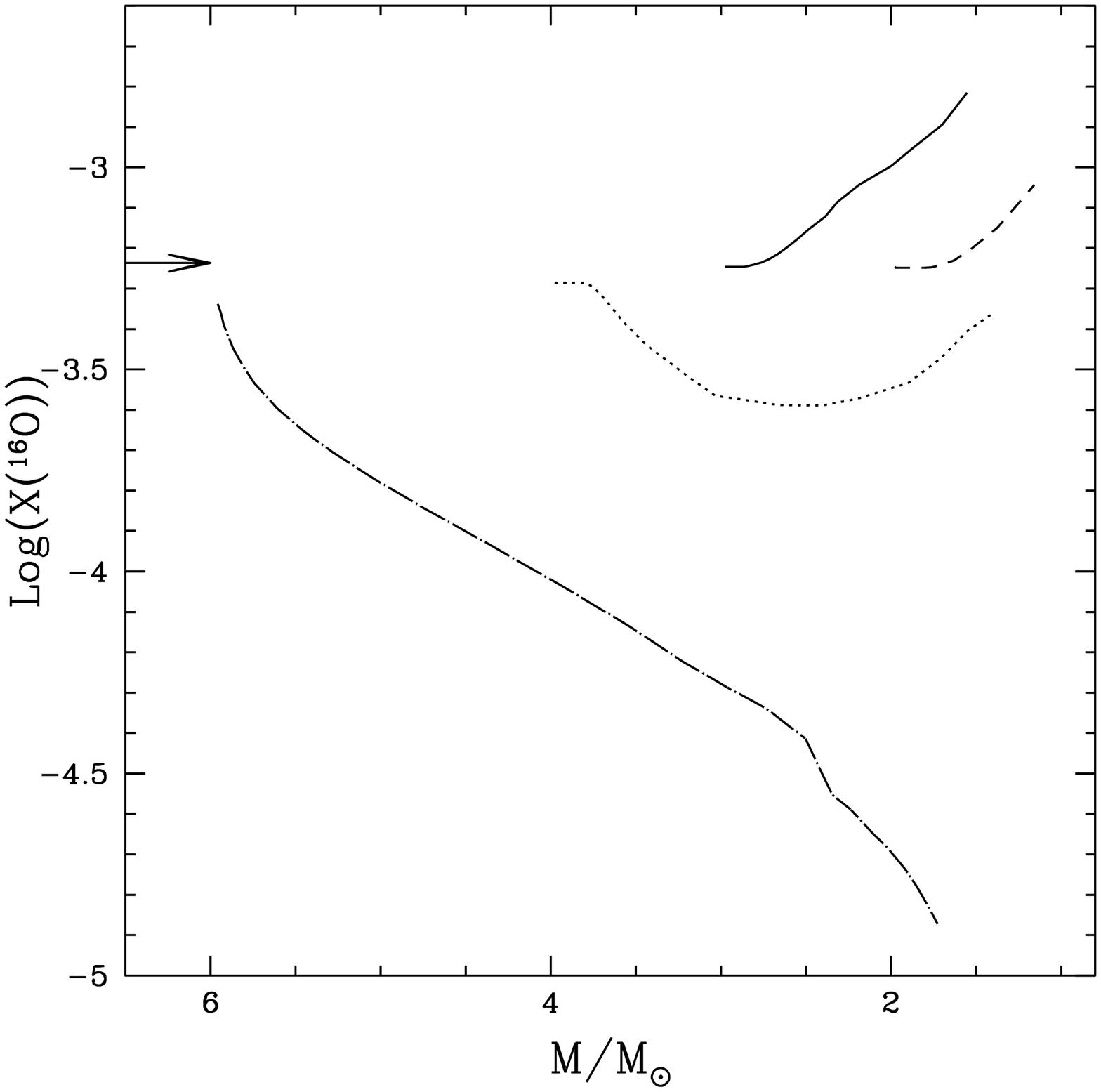}}
\end{minipage}
\begin{minipage}{0.33\textwidth}
\resizebox{1.\hsize}{!}{\includegraphics{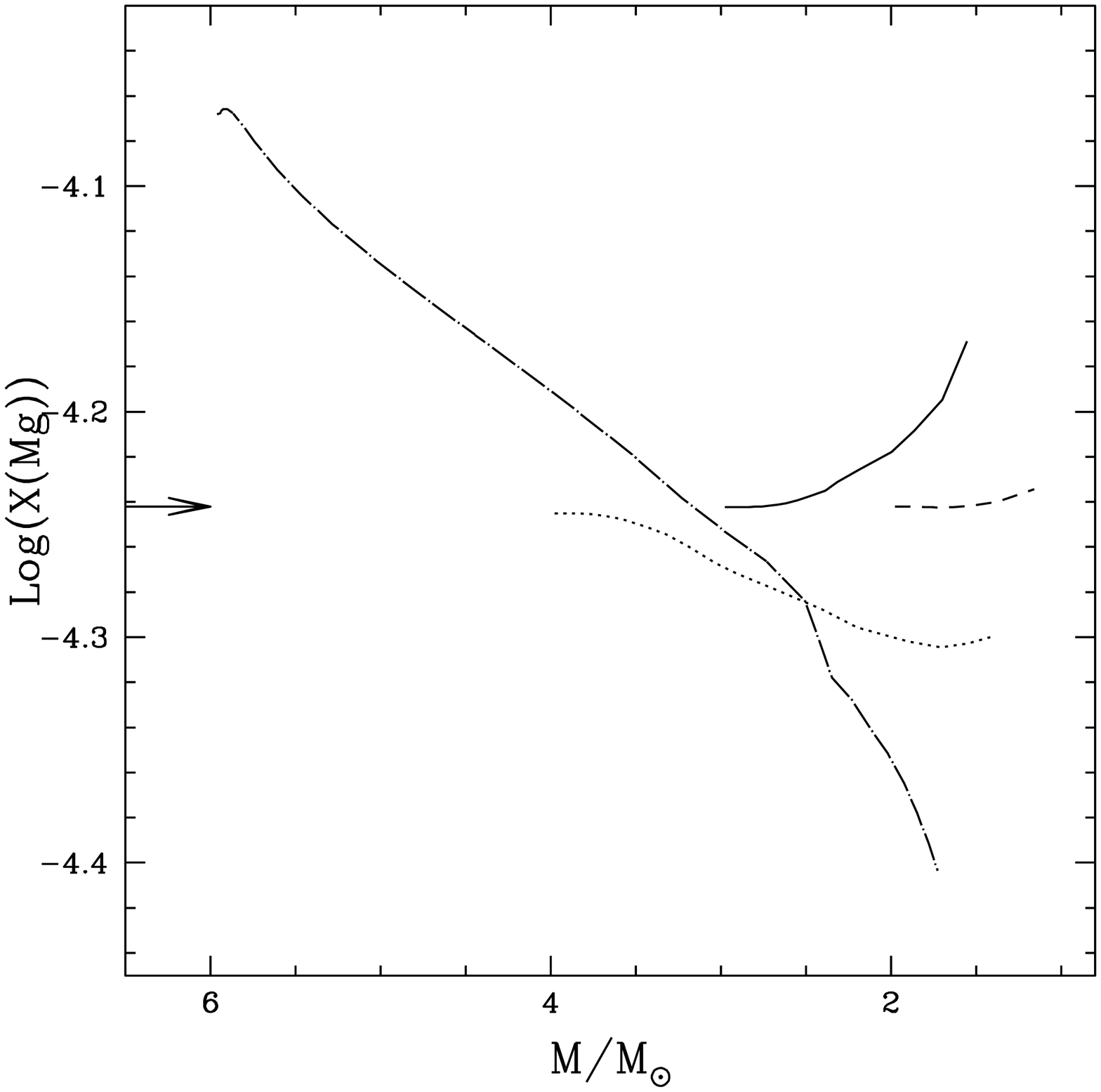}}
\end{minipage}
\caption{The evolution of the surface chemistry of models with different initial masses
(same symbols as in Fig.~\ref{figfis}) as a function of the stellar mass. The three
panels refer to the surface abundances of Carbon (Left), Oxygen (Middle) and total
Magnesium (Right). The arrows indicate the initial abundances of the same elements in 
the original mixture. We note the transition from Hot Bottom Burning, operating in the
high--mass regime, and determining a depletion of all the three elements, to the
effects of the III Dredge--up, outlined by the increase in the surface carbon mass 
fraction, dominating in the lowest masses.
}
\label{figchim}
\end{figure*}

For C--rich models, a further source of uncertainty is the adopted low--T radiative
opacities. When the C/O ratio approaches unity, the main absorbers of radiation switch from
the oxygen bearing molecules, such as H$_2$O, VO, TiO, to C--bearing molecules, such as CN,
C$_2$, C$_3$, C$_2$H$_2$, which are much more sensitive to the surface mass fractions of the
main elements: this makes the C--rich models much more sensitive to the details of the
variation of the surface chemical composition of the star. The increase in the absorption of
radiation favoured by a larger carbon content is commonly neglected in 
the computations of the AGB phase, that assume the opacity to depend only on the initial
metal content, and disregard any modification of the surface chemistry. As initially suggested by 
\citet{marigo02}, and later confirmed on the basis of full AGB computations \citep{cristallo08,
vm09,vm10}, the usage of the correct opacities favours a sudden expansion of the structure once
C/O reaches unity, which is accompanied by an increase in
the mass loss rate experienced; the consumption of the envelope is consequently faster, which
prevents the possibility that enormous amounts of carbon are accumulated in the surface regions.
An interesting consequence, to be explored also within the context of the present investigation, 
is that the general cooling of the structure that follows the approach of the C--star stage 
may potentially quench the Hot Bottom Burning, and favour a C--rich environment. \\

\begin{figure*}
\begin{minipage}{0.45\textwidth}
\resizebox{1.\hsize}{!}{\includegraphics{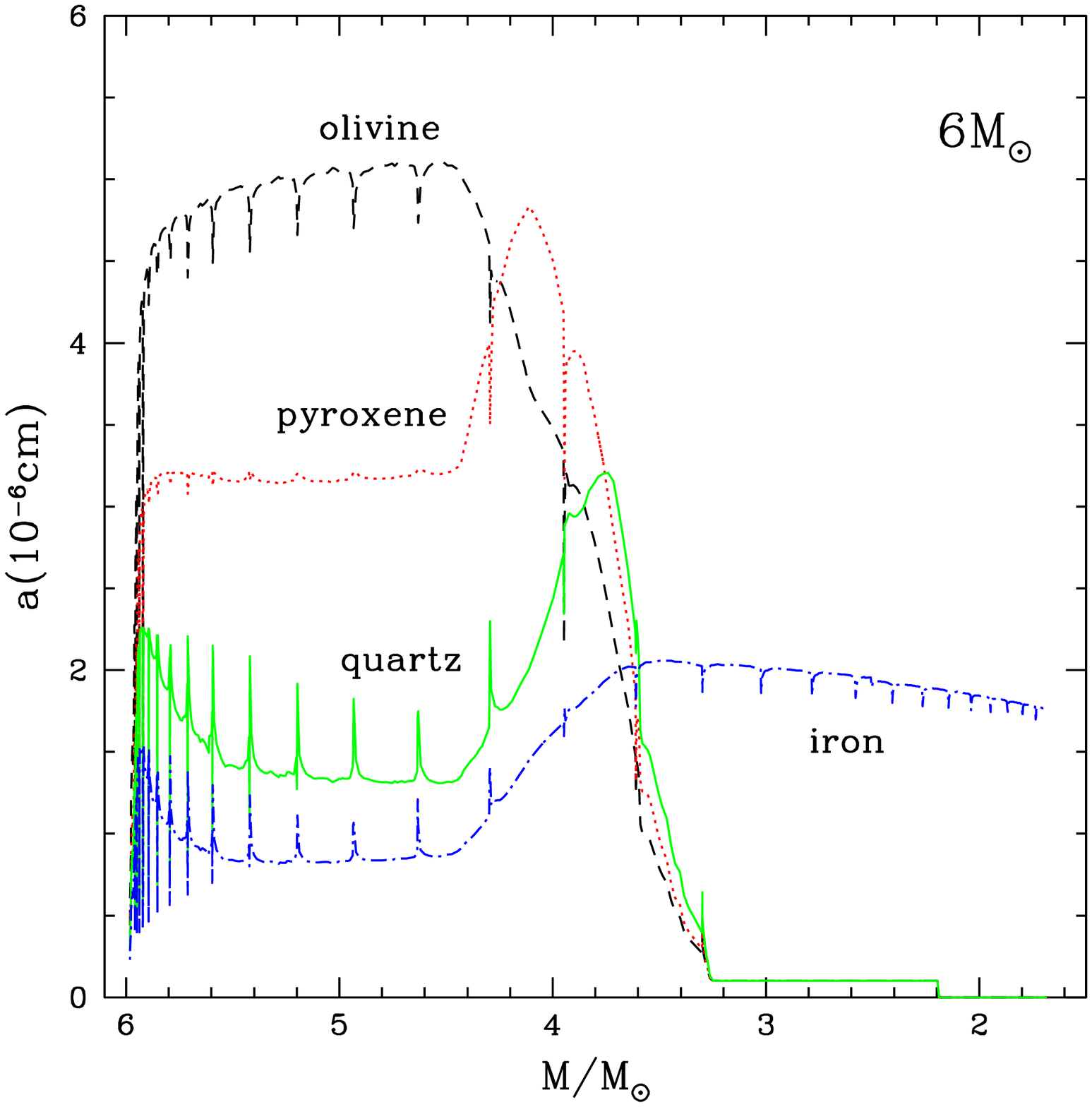}}
\end{minipage}
\begin{minipage}{0.45\textwidth}
\resizebox{1.\hsize}{!}{\includegraphics{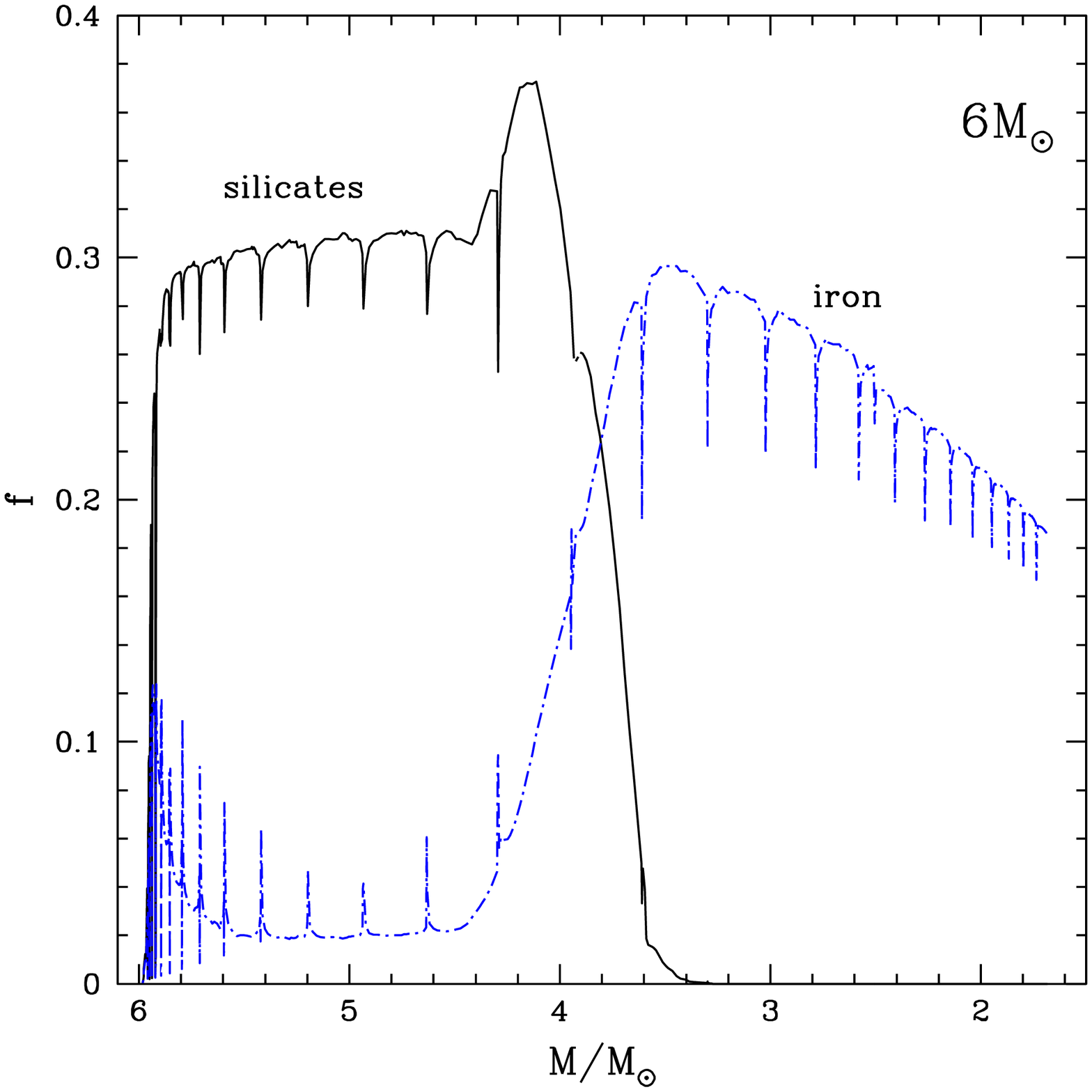}}
\end{minipage}
\caption{Left: Grain size evolution as a function of the total mass of the star during
the  AGB evolution of a model with initial mass 6M$_{\odot}$. The different lines correspond
to olivine (dashed, black), pyroxene (dotted, red), quartz (solid, green) and iron (dot--dashed,
blue). Right: The corresponding evolution of the fraction of silicon condensed into silicate--type
dust (solid, black), and of the iron condensed (dot--dashed, blue)
}
\label{fig6}
\end{figure*}

\begin{figure*}
\begin{minipage}{0.45\textwidth}
\resizebox{1.\hsize}{!}{\includegraphics{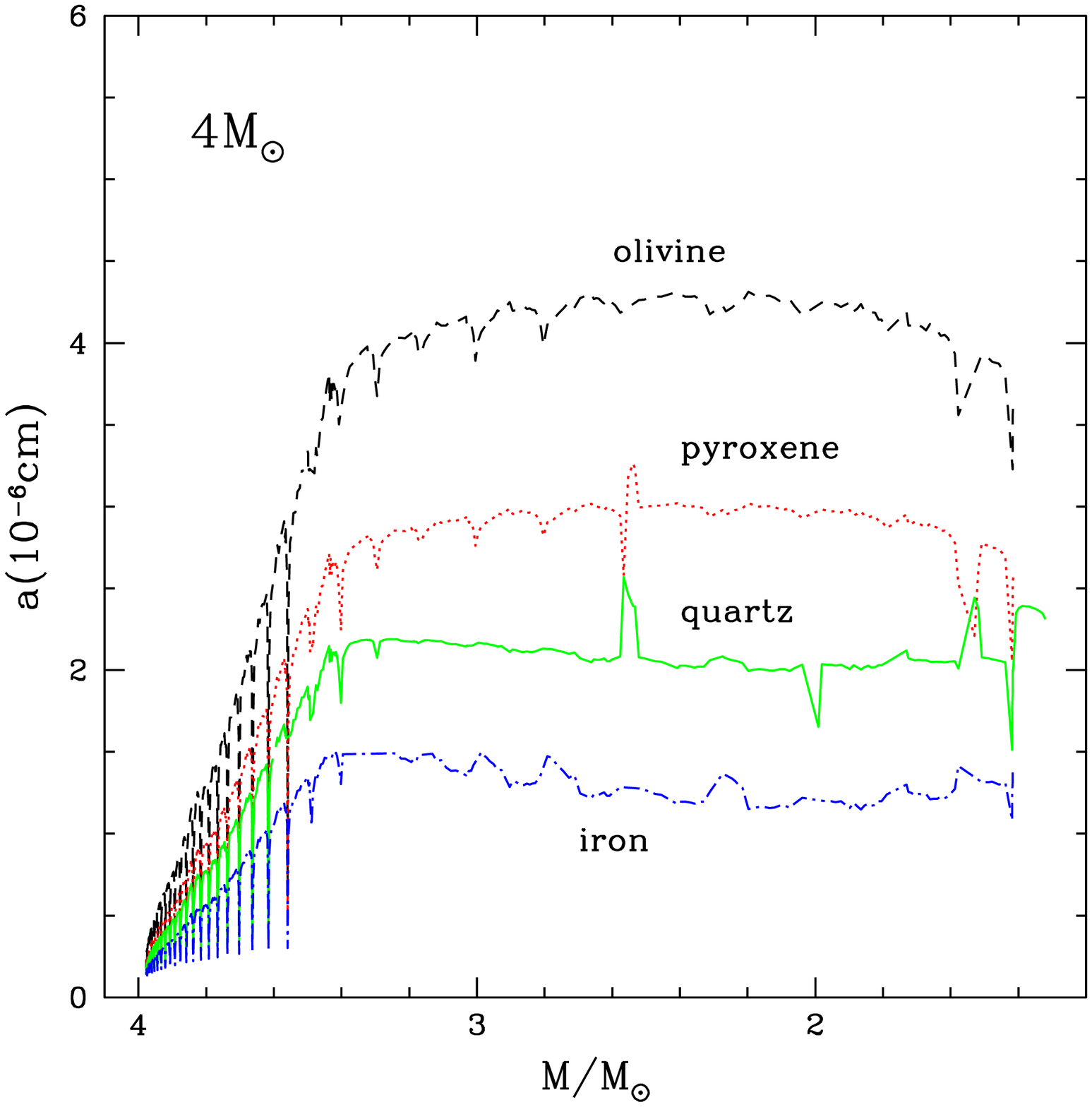}}
\end{minipage}
\begin{minipage}{0.45\textwidth}
\resizebox{1.\hsize}{!}{\includegraphics{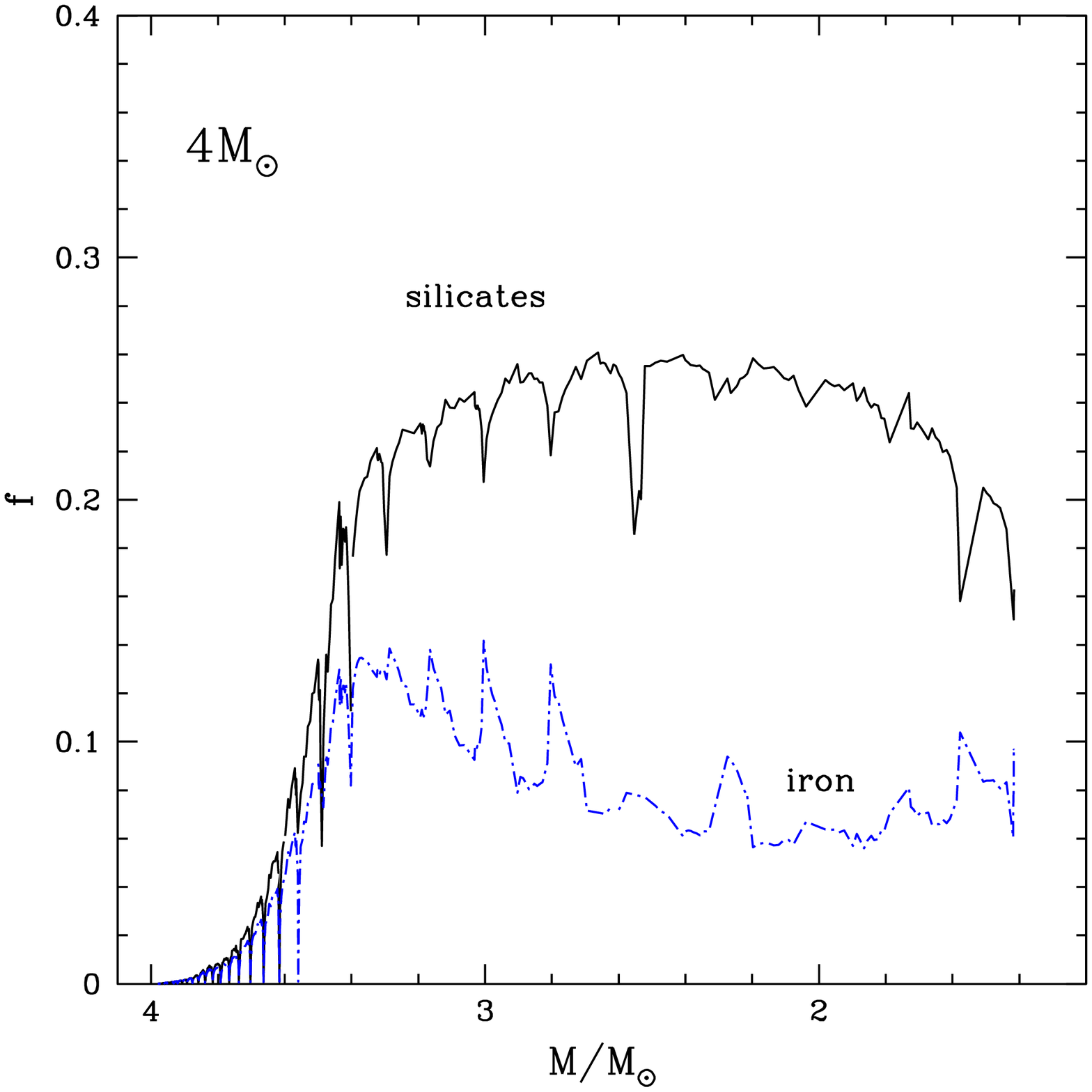}}
\end{minipage}
\caption{Left: The evolution as a function of the total mass of the star during
the  AGB evolution of the grain size (left) and the fraction of the key elements condensed (right) 
in a model of initial mass 4M$_{\odot}$. The meaning of the different
lines in the two panels is the same as in Fig.~\ref{fig6}.
}
\label{fig4}
\end{figure*}

\subsection{Hot Bottom Burning}
The description of the Hot Bottom Burning (HBB) phenomenology requires 
a full integration of the whole stellar structure, and a diffusive description of the
coupling of nuclear burning and mixing of chemicals within the envelope 
\citep{sackmann74, ventura00}. The HBB phenomenon
is a consequence of the delicate coupling between the outer region of the degenerate core, the 
CNO burning layer, and the innermost regions of the surface convective zone, and by definition
cannot be described within the framework of a synthetic modelling of the AGB phase.

Unlike the TDU, HBB conditions are reached during the interpulse phase. 
The temperature at the base of the external mantle may exceed $\sim 40\times 10^6$K, with the 
consequent activation of
p--capture nucleosynthesis. The products of such nuclear activity, due to the rapidity of the 
convective currents, are immediately transported to the surface. A mild activation of HBB
leads to the depletion of carbon, whereas a stronger HBB may potentially lead to the destruction
of the surface oxygen (T$> 70\times 10^6$K); in both cases, a great enhancement of the surface nitrogen is achieved.
The modelling of the HBB phenomenology is extremely uncertain, because the
thermodynamic stratification of the internal region of the surface convective zone is
critically dependent on the convective model adopted: a higher efficiency convective description
favours HBB \citep{renzini81}. \citet{blocker91} showed that HBB ignition determines a steep
increase in the luminosity of the star, and a significant deviation from the classic relationship
between core mass and luminosity by \citet{paczynski}. The major impact of the HBB phenomenology
on the AGB evolution was investigated by \citet{vd05a}: when convection 
is modelled following the FST scheme, such as in the reference version of the ATON code, 
HBB is found in all models for stars more massive than $\sim 3M_{\odot}$, and is accompanied by 
a rapid increase in the mass loss rate experienced by the star, and thus by a faster consumption 
of the whole envelope. Only a few thermal pulses are experienced, which prevents the possibility 
that the C--star stage is reached.

\subsection{SAGB stars}
The main physical properties of massive stars with M$_{\rm up}<$M$<$M$_{\rm ccSN}$, 
which experience the SAGB phase, have been described in detail by \citet{siess06}. 
The recent full exploration by \citet{siess10} provides 
yields that allow to estimate how these stars pollute their surroundings, 
and to infer the extent of the nucleosynthesis 
experienced at the bottom of their surface convective zone. The recent investigation
by \citet{vd11}, via a detailed comparison with the results presented in \citet{siess10},
showed that: ({\it i}) the core masses of this class of objects is so high that 
HBB conditions are easily achieved, independently of the convection modelling; 
({\it ii}) the strength of the thermal pulses experienced is modest, thus TDU never occurs,
and the C--star stage is never achieved; ({\it iii}) the mass loss treatment plays a major role 
in determining the physical evolution of SAGBs. 
When a treatment a la \citet{VW93} is adopted, the star experiences a large number 
of TPs; the extent of the nucleosynthesis at the bottom of the convective mantle is so high
that oxygen is severely depleted \citep{siess10}. On the contrary, when the \citet{blocker95}
recipe is adopted, the star lose all the envelope before a strong nucleosynthesis may occur.

The SAGB stars are thus expected to experience a strong mass loss since the beginning of their
TP evolution, and to show-up traces of p--capture nucleosynthesis, although the rapid
consumption of their external mantle prevents the achievement of extended modifications of
their surface chemistry.

\section{Which kind of dust from AGB and SAGB stars?}
\label{sec:results}
The quantity of dust formed around a star depends on the gas density in the region where
condensation occurs. Within the present schematization, mass conservation imposes a direct
relationship between $\dot M$ and $\rho$ (see Eq.~5): higher densities correspond to
large mass loss rates, which renders AGB and SAGB stars particularly suitable for the dust
formation process in their surroundings. In the following we describe the results of the model 
for stars with initial masses in the range $1 M_{\odot} \le M \le 8 M_{\odot}$ and initial 
metallicity of $Z=0.001$. The main evolutionary properties of these models were presented and 
discussed in \citet{vm10} and \citet{vd11}. Fig.~\ref{figfis} shows the evolution of the main
physical properties of four models differing in their initial mass; the three panels refer to the
whole AGB phase, during which the stellar envelope is gradually consumed, with a decrease in the
stellar mass. The variation of the surface chemistry in the same models is shown in 
Fig.~\ref{figchim}. We limit this description to carbon, oxygen and magnesium, because iron
is unchanged, and silicon shows a modest variation ($\sim +0.05$ dex), and only in the
6M$_{\odot}$model.

\subsection{Dust formed under HBB conditions}
The left panels of Figs.~\ref{fig6} and ~\ref{fig4} shows the evolution of dust grain 
sizes formed in winds of stars with masses $M = 6 M_{\odot}$ and $M = 4 M_{\odot}$ 
(right panel), that experience HBB. The right panels show the fraction of the key
elements condensed into dust: silicon for olivine, pyroxene and quartz, and iron
for solid iron. The three panels of 
Fig.~\ref{figfis} show the evolution of their main physical quantities during the AGB phase
(dashed-dotted and dotted lines, for 6M$_{\odot}$ and 4M$_{\odot}$, respectively),
and demonstrate the large luminosities and mass loss rates reached, in particular by the 
$M = 6 M_{\odot}$ model.

The depletion of the surface carbon (indicated by the initial drop in the carbon mass
fraction in the left panel of Fig.~\ref{figchim}) prevents the C--star stage to be reached, 
thus allowing only the formation of silicates and solid iron. Instead of the time coordinate, 
which increases from left to right, in both Figs.~\ref{fig6} and ~\ref{fig4} we show the results 
as a function of the mass of the star: this helps a better understanding of the amount of dust 
produced, because the latter depends on the mass lost by the star when the size of the grains 
is the highest.

Given our choice of convective overshoot from the core during the H--burning phase, 
the left panel represents the evolution of the largest mass model 
which evolves through the AGB phase. We can distinguish different phases during the
evolution. Dust begins to form as soon as HBB begins, and the mass loss
increases. Olivine is the most abundant species and it forms close to the stellar surface; 
as HBB becomes stronger, the mass loss rate further increases, leading to 
an increase in the amount of olivine formed but, at the same time, causing a decrease 
of the least abundant species, i.e. quartz and iron. This result, in agreement 
with the analysis of \citet{fg01}, is due to the rapid acceleration experienced by the 
wind as soon as the first species form, which decreases the density of the key species.

The size of the olivine and pyroxene grains decreases at each thermal pulse, as a 
consequence of the temporary drop in the luminosity (hence of the mass loss rate) of the star.
Quartz and iron follow the opposite behaviour, because the smaller mass loss reduces
the differences among the amount of dust formed for each species.

We see from Fig.~\ref{fig6} that in the $6M_{\odot}$ model, when the total mass of the star 
drops below $\sim 4M_{\odot}$, the depletion of surface oxygen (clearly indicated by the slope
of the dashed--dotted line in the middle panel of Fig.~\ref{figchim}) determines a scarcity of water 
molecules, required for the formation of the three species of silicates considered (see Table 1). 
Because the formation of a molecule of olivine requires three H$_2$O molecules, 
olivine production stops, followed by pyroxene (which needs only two water molecules), 
and quartz (one water molecule). From this point on, only iron grains are produced. 

We thus find that massive AGB models that experience strong HBB conditions produce a large 
amount of silicate--type dust during the first part of their AGB evolution, and end--up with 
iron dust production once the surface abundance of oxygen is strongly reduced. 
A word of caution concerning this latter possibility is needed here, because the occurrence of
an iron--driven outflow might be an artificial effect of the grey description adopted in
the present modelling, which favours smaller condensation distances for the iron grains.
Simple analytical estimates \citep{hofner09}, supported by detailed models \citep{Woitke},
demonstrated that the wavelength dependence of the corresponding grain opacities in the
near--IR leads to condensation distances well beyond the reach of shock waves, which means
that iron cannot start an outflow, unless the wind is triggered by other types of dust. The
possibility of an outflow driven only by iron is thus highly uncertain.

The $4M_{\odot}$ model (see Fig.~\ref{fig4}) evolves at smaller core masses, and can be taken 
as representative of a star experiencing a milder HBB.
Compared to the 6M$_{\odot}$ case, due to the smaller $\dot M$, less silicates are produced
in the early AGB phases; on the other hand, the depletion of oxygen is less severe (see middle
panel of Fig.~\ref{figchim}), thus silicates continue to form throughout the AGB evolution. 
These two effects compensate each other, thus the two models, at the end of the evolution, produce 
approximately the same quantity of silicates and a total dust mass of 
$\sim (1.2-2) \times 10^{-4} M_{\odot}$ (see Table 2). The lower mass loss also leads to smaller 
differences among the various types of dust formed.

\begin{figure*}
\begin{minipage}{0.45\textwidth}
\resizebox{1.\hsize}{!}{\includegraphics{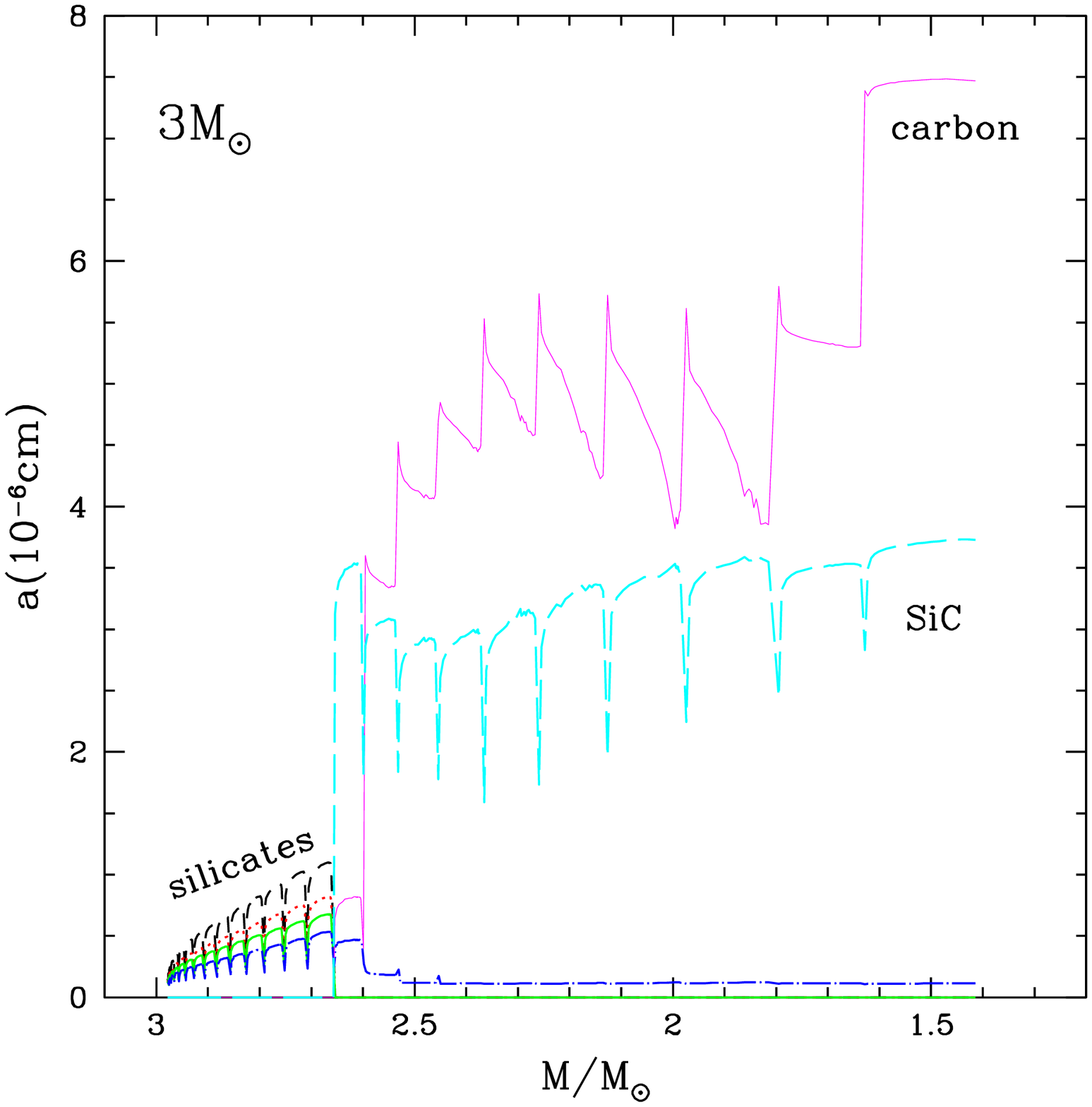}}
\end{minipage}
\begin{minipage}{0.45\textwidth}
\resizebox{1.\hsize}{!}{\includegraphics{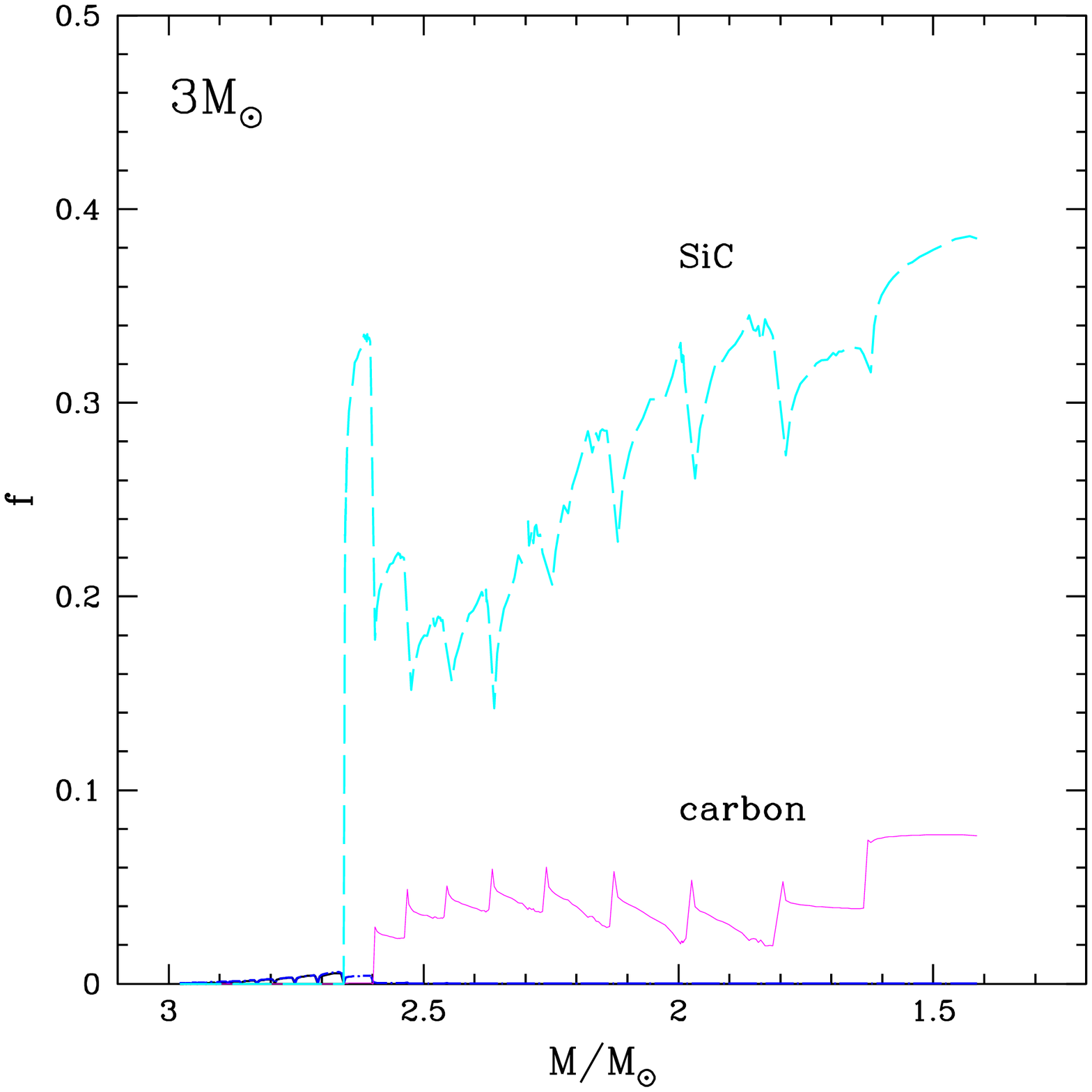}}
\end{minipage}
\caption{The same as Fig.~\ref{fig6}, but for a stellar model with
initial mass 3M$_{\odot}$. Line coding for silicates and iron (whose production is
however negligible in the present model) is the same as in Fig.~\ref{fig6}. 
Light solid, magenta lines indicate the evolution of carbon grain size (left panel) and
of the fraction of carbon condensed (right); Long--dashed, cyan lines show the variation
of the SiC grain size (left) and of the fraction of silicon condensed into SiC (right). 
We note the effects of the transition from M stars to C stars, when no 
silicates are produced.
}
\label{fig3}
\end{figure*}

\begin{figure*}
\begin{minipage}{0.45\textwidth}
\resizebox{1.\hsize}{!}{\includegraphics{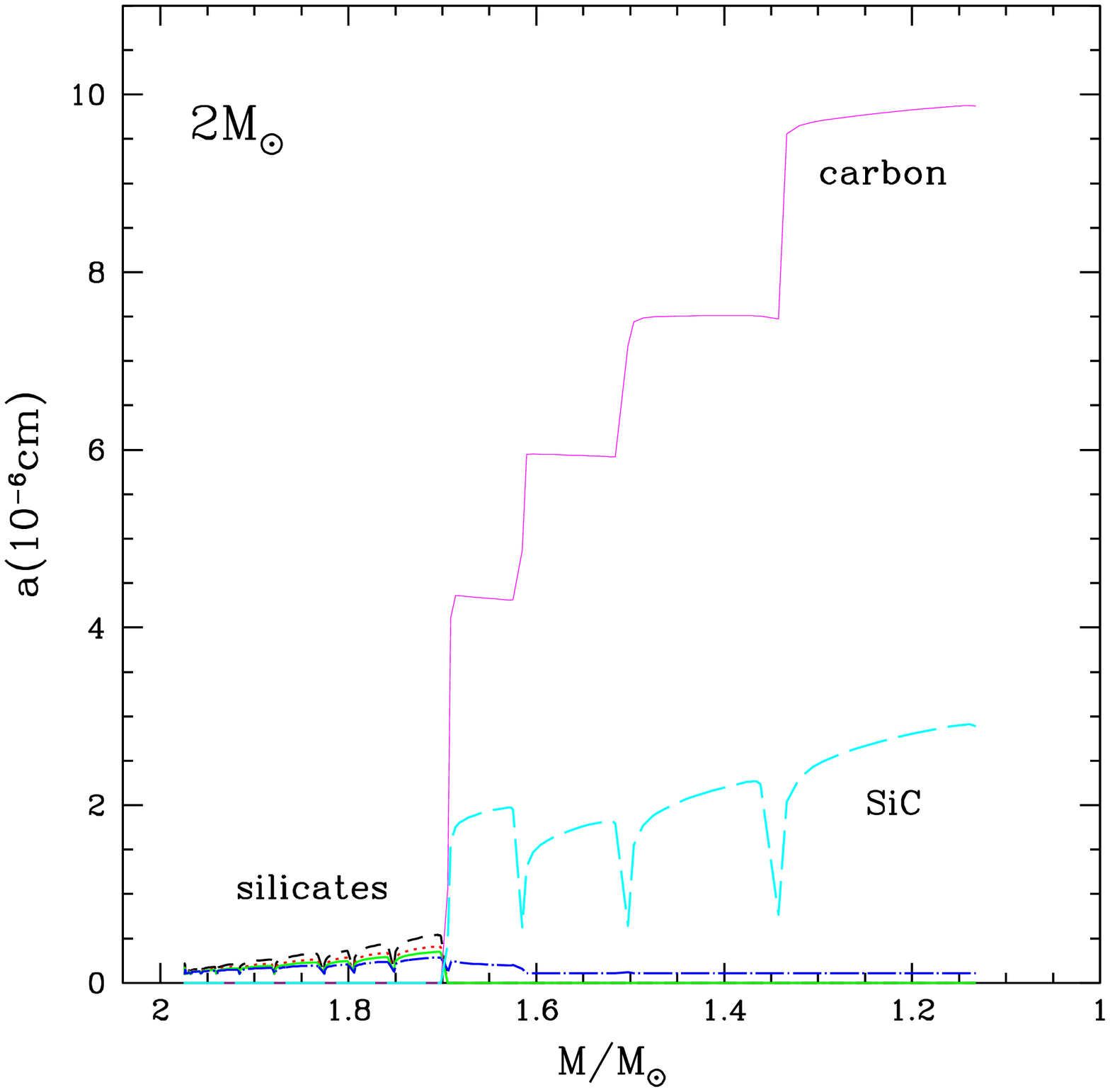}}
\end{minipage}
\begin{minipage}{0.45\textwidth}
\resizebox{1.\hsize}{!}{\includegraphics{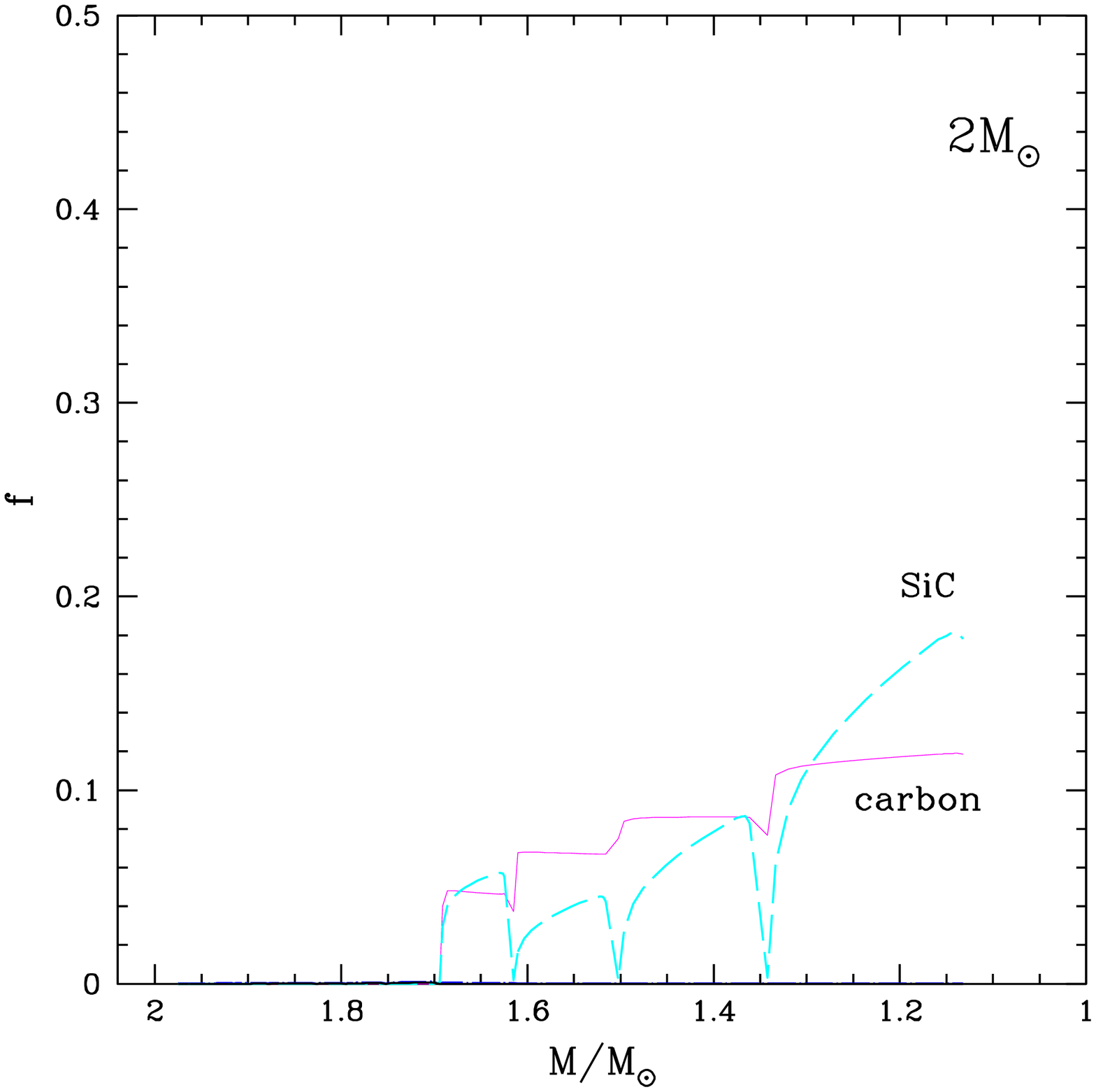}}
\end{minipage}
\caption{The same as Fig.~\ref{fig3}, but for a stellar model with
initial 2M$_{\odot}$ (right). 
}
\label{fig2}
\end{figure*}

\subsection{Dust from C--rich environments}
Models whose initial mass is below 3.5$M_{\odot}$ hardly reach HBB conditions, and evolve
at smaller luminosities compared to their more massive counterparts (see the left panel of
Fig.~\ref{figfis}). The repeated TDU episodes eventually leads to the formation of a C--star,
as indicated by the great increase in the surface carbon content for the models with
initial mass 2M$_{\odot}$ and 3M$_{\odot}$ shown in the left panel of Fig.~\ref{figchim}.
This is accompanied by a
switch from silicate--type dust to C--rich particles. In the left panels of 
Figs.~\ref{fig3} and \ref{fig2} we show the grain size evolution as a function of the mass 
of the star for two models with $3M_{\odot}$ and $2 M_{\odot}$; the right panels show the
degree of condensation of the key elements (essentially, carbon and silicon in this case).
In both cases we note an initial modest production of silicates, that proceeds until C/O 
reaches unity. From that point on, solid carbon and
SiC are formed. The decrement in the dimension of carbon grains during each
interpulse period in the 3$M_{\odot}$ model is an effect of HBB, which is completely
absent in the 2$M_{\odot}$ model. In this latter case, representative of all the models 
that show no effects of HBB, a large amount of carbon grains are expected to form around 
the star. Unlike their more massive counterparts, low--mass AGBs progressively increase their 
surface carbon content due to the repeated TDUs, and produce more and more dust as they evolve, 
until no mass is left within the convective envelope. The total mass of dust formed by stars 
with $M \le 3.5 M_{\odot}$ ranges between $\sim (0.01-7.7) \times 10^{-4} M_{\odot}$ (see 
Table 2).

\subsection{Dust from SAGB stars}
\begin{figure}
\resizebox{1.\hsize}{!}{\includegraphics{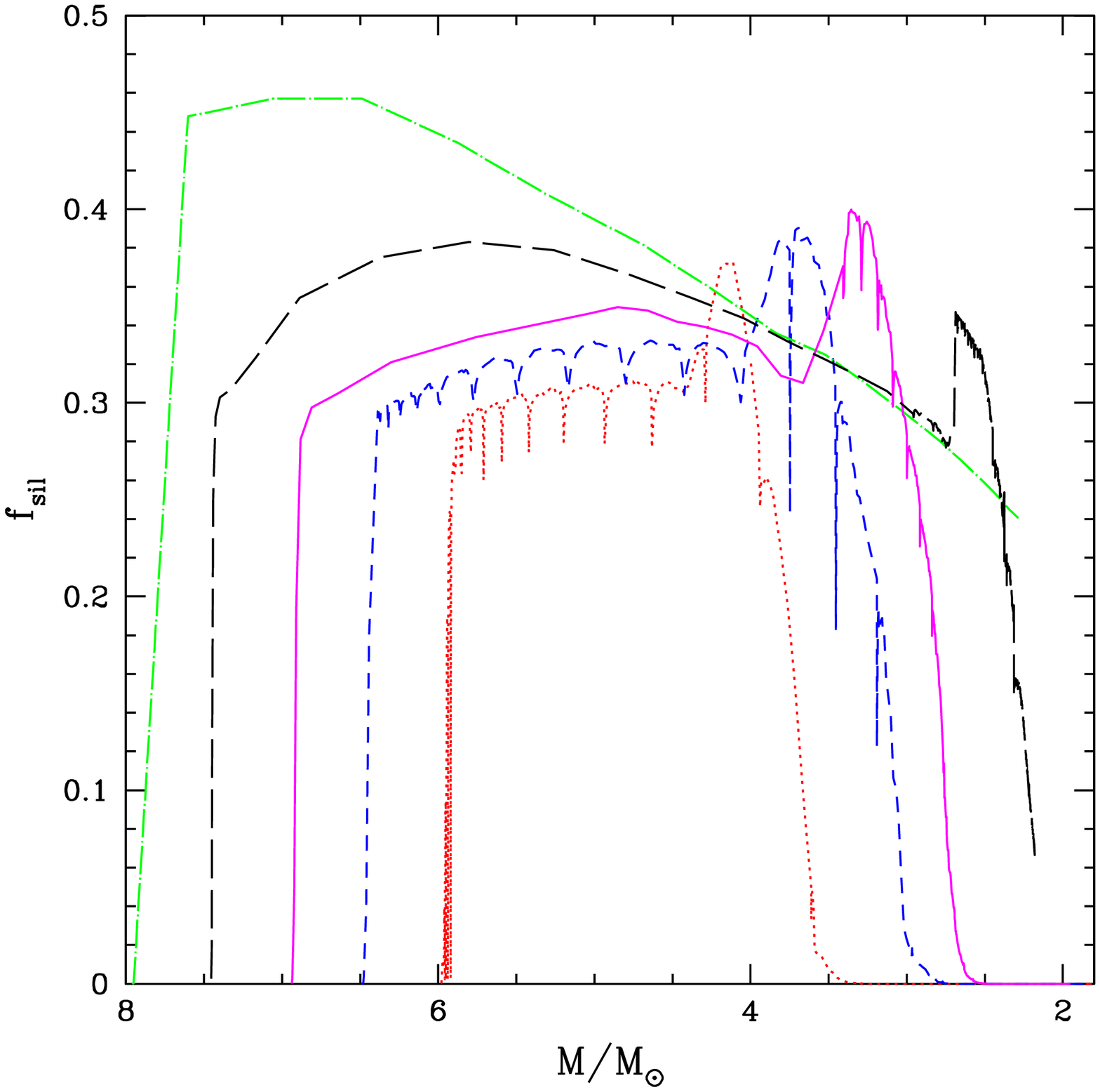}}
\caption{The total fraction of silicon condensed into silicate dust grains as a function of the
total stellar mass in SAGB models with initial masses of 6.5 (dashed, blue), 7 (solid, magenta), 
7.5 (long--dashed, black), and 8 $M_{\odot}$ (dot--dashed, green). For comparison, we also show 
the results obtained for the most massive AGB model with $M = 6 M_{\odot}$ (dotted, red line).
}
\label{figsagb}
\end{figure}

SAGB stars evolve at large luminosities since the first TPs, and experience a strong
mass loss rate that favours an early consumption of their convective envelope.
Fig.~\ref{figsagb} shows the variation of the overall fraction of silicon condensed
into silicate--type dust as a function of the total stellar mass. The different lines
correspond to SAGB models with masses in the range 
$6.5 M_{\odot} \leq M \leq 8 M_{\odot}$ published in \citet{vd11}. 
For comparison, the most massive AGB model ($6 M_{\odot}$) previously discussed is also shown. 
It is evident that the mass of silicate dust grains produced by SAGBs is larger than in the 
$6 M_{\odot}$ model, for two main reasons: ({\it i}) since SAGBs evolve at large core masses, 
their luminosity and mass loss are
extraordinarily large since the very first TPs and ({\it ii}) unlike the most massive AGBs, 
oxygen is not depleted extensively, because the envelope is lost before the nucleosynthesis 
experienced at the bottom of their surface mantle can reach very advanced stages. Thus, 
silicates continue to form during the whole TPs evolution because water molecules are 
always available in their surroundings. The total masses of dust formed by SAGBs vary in the
range $\sim (3 - 7)\times 10^{-4} M_{\odot}$ (see Table 2). 

\subsection{Stardust from AGB and SAGB stars}
In Table 2 we list the total dust masses and the masses of individual dust species
predicted by the model for the investigated grid of stellar masses. Similarly
to other important AGB phenomena, the key quantity which determines the composition of dust 
formed around the star is the core mass, $M_{\rm core}$: models with small $M_{\rm core}$ 
achieve only a modest production of silicates, whereas carbon is produced in great quantities.
In the more massive objects the opposite situation occurs, since HBB prevents the
formation of any carbon--rich dust species, whereas a strong production of silicates is favoured.

\begin{figure}
\resizebox{1.\hsize}{!}{\includegraphics{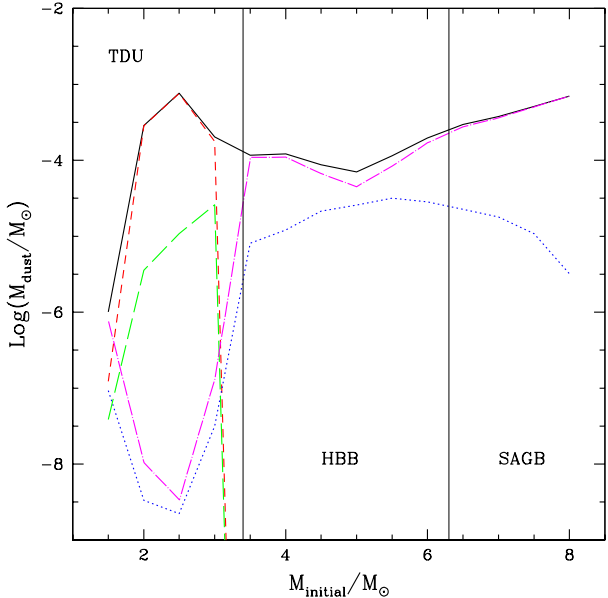}}
\caption{The total mass of dust produced by AGB \& SAGB models of different masses (solid,
black line). We also show the total mass of the individual components, indicated as
follows: Silicates (dot--dashed, magenta), carbon (dashed, red), SiC (long--dashed, green),
iron (dotted, blue).
}
\label{fguess}
\end{figure}

Fig.~\ref{fguess} shows the total mass of dust as a function of the initial stellar
mass (solid line) and the contribution of individual dust species (i.e. silicates, 
iron, carbon, and SiC). The results refer to AGB and SAGB models calculated with the
reference ATON model discussed in Sect.\ref{sec:results}. 

We note a doubled--peaked distribution, with a maximum production of carbon by low--mass AGBs and 
of silicates by SAGBs; stars with 
$3M_{\odot} < M < 5M_{\odot}$ provide the smallest contribution
to dust formation, because the production of carbon is inhibited by HBB, which
is however too soft to allow a large silicates production.
At stellar masses around $\sim 3M_{\odot}$ 
(this result depends on the treatment of the low--T molecular opacities, 
see Sec.~\ref{sec:lowTopacities}) we note the transition from 
carbon--rich to silicate--rich dust, which is a pure effect of HBB, 
prevailing over the TDU effects.

The mass of silicates formed by stars in the mass range $[4-6]M_{\odot}$ is approximately constant: 
this is a consequence of the strong production of silicates, achieved in the early AGB phases 
in the more massive models, counterbalanced by the scarcity of the water molecules 
available for the condensation process, due to the very strong and fast depletion of 
oxygen, that characterize the latest evolutionary stages. 
The SAGB models, are, under the present assumptions, the most efficient silicates producers, 
for the reasons discussed in Sect.~4.3.

\section{Dependence of dust production on physical parameters}

In the previous section we have presented the results of the dust formation model
for different stellar masses. Here we discuss how uncertainties in the modelling
of the AGB/SAGB evolution and in the physics of dust formation may affect 
the robustness of these findings.
In particular, we examine the role played by the adopted low--T molecular opacities, 
the mass loss treatment, and the chosen set of sticking coefficients. 
A discussion of the impact of the adopted convective model is presented in Sect. 6.

\subsection{Low--T molecular opacities}
\label{sec:lowTopacities}
The models presented and discussed so far (see, in particular, Figs.~\ref{fig3} and \ref{fig2}) 
are based on a modern and updated method to determine the
molecular opacities in the low temperature regime: the AESOPUS tool \citep{marigo09}, 
that accounts for the changes in the chemistry of the envelope to calculate the absorption 
coefficient, $k$, in the most external regions of the star.

In the traditional approach, the evolution of the surface abundances of the various species
is ignored: the opacity is determined on the basis of the initial composition of the star, 
neglecting the increase in the surface carbon abundance provided by TDU.

Since previous studies of dust production by AGBs 
rely on this latter approximation, we have quantified the differences introduced by the new 
method to calculate the opacities in terms of the total mass of dust formed. 
We thus calculate a further set of AGB models in which
the traditional method was used to compute $k$ in the low--T regime.

As long as the C/O ratio is $< 1$, such as in the most massive AGB models which experience HBB, 
the evolution is independent of the opacity treatment \citep{vm10}. We thus restrict our 
analysis to lower mass stars (M$\leq 3M_{\odot}$), whose surface chemistry is dominated by 
TDU.
\begin{figure}
\resizebox{1.\hsize}{!}{\includegraphics{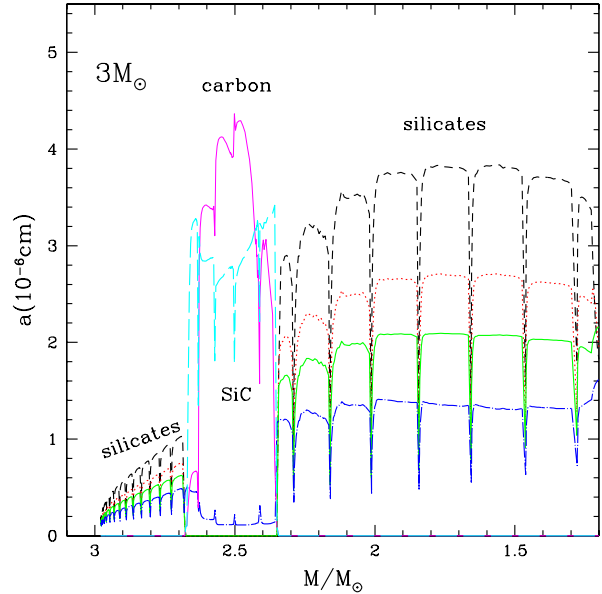}}
\caption{The dust produced by a 3M$_{\odot}$ model in which, as in previous works, 
the increase in the surface carbon is neglected in the computation of the low--T opacities. 
The dimensions of the 
different types of dust produced are indicated with the same symbols as in Fig.~\ref{fig32}.
The comparison with the results shown in the left panel of Fig.~\ref{fig32} allows us to
understand the role played by the low--T opacity description: we note that in this case
HBB is not extinguished, thus the production of carbon--rich dust is limited to a narrow
time--interval, before strong HBB destroys the surface carbon available.
}
\label{figk}
\end{figure}

Fig.~\ref{figk} shows the resulting grain size evolution for a $3 M_{\odot}$ star; 
this is to be compared with the left panel of Fig.~\ref{fig3}, that refers to a model with the
same mass, calculated with the AESOPUS tool. Both models achieve an early small production 
of silicate--type dust, that proceeds until repeated TDU episodes allow to reach the 
C--star stage. From this point on, the predictions diverge. When the opacities are computed
with the AESOPUS tool, carbon--rich dust continues to be produced during the whole AGB phase. 
On the other hand (see Fig.~\ref{figk}), when the traditional opacities are used, 
C and SiC grains are produced only within a narrow range of stellar mass, 
after which silicates production takes over again. 
This is a consequence of the quenching of HBB, that was early predicted by 
\citet{marigo02}, and later confirmed by \citet{vm09}: because the core 
mass of a $3 M_{\odot}$ star is the minimum allowing HBB conditions, the expansion of the
envelope following the increase in the surface carbon is sufficient to
turn off HBB:  this is favoured by the rapid
increase of AESOPUS opacities that accompanies the achievement of the 
C--star stage. The model shown in Fig.~\ref{figk} does not follow this behaviour
due to the use of classical opacities; the star eventually reaches HBB conditions
and silicates production takes over again.

The 3M$_{\odot}$ model just discussed lies within the narrow range of masses 
($\delta M\sim 0.5M_{\odot}$) where the differences caused by the two opacity 
treatments are most dramatic, affecting both the resulting grain composition and
the total mass of dust produced: when the traditional opacitities are adopted, the
total mass of dust produced is $7\times 10^{-5}M_{\odot}$ (to be compared with 
the corresponding value in Tab.~2, i.e. $\sim 2\times 10^{-4}M_{\odot}$), most of which is 
silicate--type dust. Stars with M$\leq 2.5M_{\odot}$ 
do not experience any HBB and the low--T opacity treatment affects only 
the total mass of C--rich dust formed, which is approximately a factor $\sim 2$
smaller when the AESOPUS tool is used. In fact, the increase in the surface carbon 
abundance is accompanied by a general expansion of the whole structure, and by a faster consumption 
of the whole envelope, resulting in a smaller surface carbon content.  

It is important to remark that the above conclusions depend on the adopted mass loss rate.
As shown by \citet{vm10}, the physical and chemical evolution of low--mass AGB stars 
depend on the interplay between mass loss rate and molecular opacities: 
a smaller $\dot M$ favours HBB and thus shifts downwards the range of masses dominated by TDU, 
around which the formation of carbon--type dust occurs.

\begin{table*}
\caption{Dust mass produced by AGB and SAGB models, assuming an initial metallicity
$Z=0.001$. The initial stellar mass $M$ and the core mass
$M_{\rm core}$ are reported in the first and second columns. The total mass of dust, $M_{\rm d}$ and
the mass of pyroxene ($M_{\rm py}$), quarz ($M_{\rm qu}$), solid iron ($M_{\rm ir}$), 
solid carbon ($M_{\rm C}$) and SiC ($M_{\rm SiC}$) are also shown. All the masses are expressed in solar
units.}
\label{dustmass}
\begin{tabular}{ccccccccc}
\hline
\hline
$M$ & $M_{\rm core}$  &  $M_{\rm d}$ & $M_{\rm ol}$ & $M_{\rm py}$ & 
$M_{\rm qu}$ & $M_{\rm ir}$ & $M_{\rm C}$ & $M_{\rm SiC}$  \\
\hline
1.5  &  0.530  &  1.01d-06  &  4.97d-07 &  1.89d-07 &  6.97d-08 &  9.25d-08 &  1.23d-07  & 3.85d-08 \\
2.0  &  0.533  &  2.88d-04  &  6.19d-09 &  2.88d-09 &  1.30d-09 &  3.34d-09 &  2.84d-04  & 3.55d-06 \\
2.5  &  0.645  &  7.67d-04  &  1.97d-09 &  9.68d-10 &  4.52d-10 &  2.23d-09 &  7.56d-04  & 1.08d-05 \\
3.0  &  0.711  &  2.03d-04  &  8.50d-08 &  3.50d-08 &  1.43d-08 &  3.17d-08 &  1.77d-04  & 2.60d-05 \\
3.5  &  0.812  &  1.17d-04  &  7.40d-05 &  2.63d-05 &  8.19d-06 &  8.12d-06 &    -      &  -      \\
4.0  &  0.842  &  1.21d-04  &  7.22d-05 &  2.70d-05 &  9.56d-06 &  1.21d-05 &    -      &  -      \\
4.5  &  0.877  &  8.76d-05  &  3.55d-05 &  2.12d-05 &  9.59d-06 &  2.13d-05 &    -      &  -      \\
5.0  &  0.914  &  7.05d-05  &  2.66d-05 &  1.31d-05 &  4.97d-06 &  2.58d-05 &    -      &  -      \\
5.5  &  0.959  &  1.15d-04  &  5.18d-05 &  2.53d-05 &  5.93d-06 &  3.16d-05 &    -      &  -      \\
6.0  &  1.012  &  1.96d-04  &  1.11d-04 &  5.04d-05 &  6.64d-06 &  2.84d-05 &    -      &  -      \\
6.5  &  1.086  &  2.97d-04  &  1.94d-04 &  7.29d-05 &  7.63d-06 &  2.26d-05 &    -      &  -      \\
7.0  &  1.260  &  3.77d-04  &  2.74d-04 &  7.81d-05 &  7.09d-06 &  1.79d-05 &    -      &  -      \\
7.5  &  1.308  &  5.11d-04  &  4.14d-04 &  7.33d-05 &  1.33d-05 &  1.08d-05 &    -      &  -      \\
8.0  &  1.348  &  6.98d-04  &  6.11d-04 &  8.02d-05 &  4.12d-06 &  3.21d-06 &    -      &  -      \\
\hline
\end{tabular}
\end{table*}

\subsection{The mass loss treatment}
The mass loss description has important effects on dust formation around AGBs. In fact,
the surface chemistry of the star, which determines the composition of dust species, 
results from the interplay among TDU, HBB, and the consumption rate of the envelope. 
This is particularly important for massive AGBs and SAGBs, because the degree of the 
p--capture nucleosynthesis experienced at the base of the convective zone, hence the final
elemental abundances, depends on the mass left in the envelope at each evolutionary stage. 
Therefore, a major issue is the competition between the rates of the relevant nuclear reactions
and the time--scale for envelope consumption. 
In the reference ATON model, the mass loss treatment follows the prescription by
\citet{blocker95}. To quantify the uncertainties introduced by the mass loss treatment
we have explored models where the \citet{VW93} prescription is used.

\begin{figure*}
\begin{minipage}{0.45\textwidth}
\resizebox{1.\hsize}{!}{\includegraphics{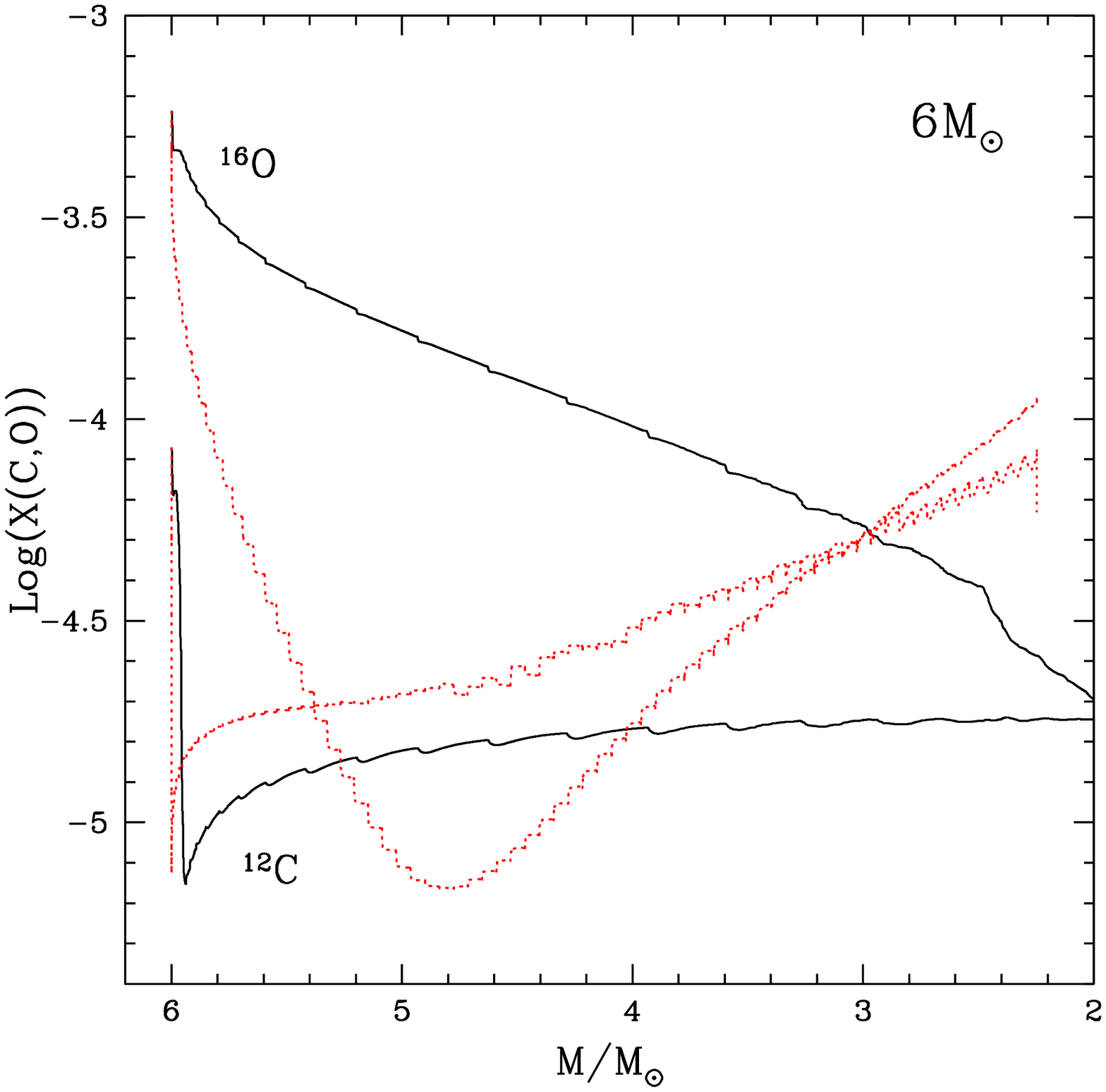}}
\end{minipage}
\begin{minipage}{0.45\textwidth}
\resizebox{1.\hsize}{!}{\includegraphics{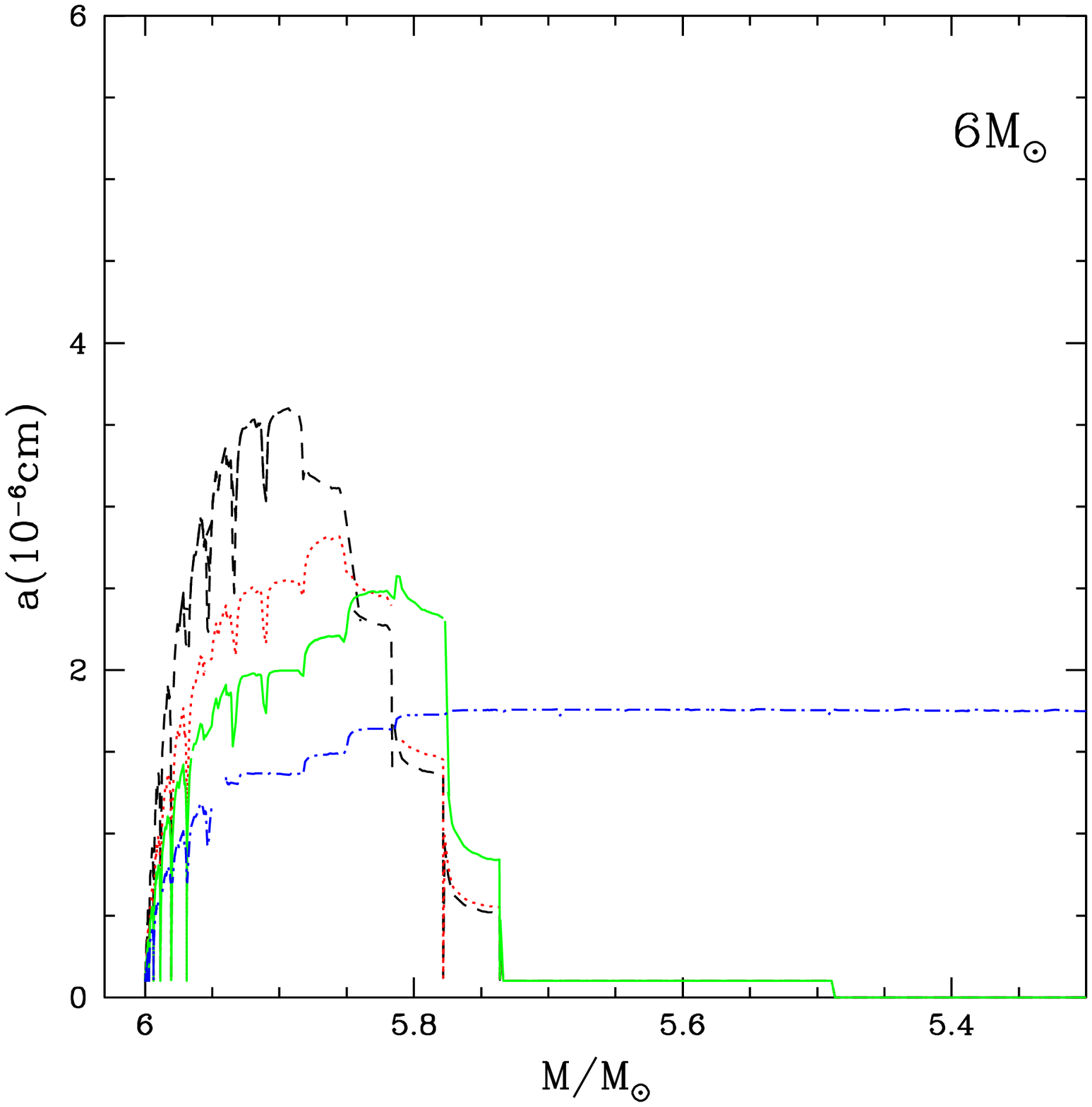}}
\end{minipage}
\caption{Left: the variaton of the surface content of oxygen and carbon in two models
of mass 6M$_{\odot}$ in which the mass loss was modelled according to the prescriptions
by \citet{blocker95} (solid, black line) and \citet{VW93} (dotted, red). In this latter case
only a little fraction of the stellar mass is lost by the time that the surface oxygen
almost disappears, thus the star enters the C--star regime, and the silicates production stops.
The right panel (to be compared to the left panel of Fig.~\ref{fig64}, that shows the
corresponding evolution calculated with the \citet{blocker95} recipe for mass loss) shows that
the silicates production is extremely poor, and the star produces iron for the almost
totality of its evolution. Note that the small carbon abundances inhibit the formation of
C--rich dust particles, even in conditions of C/O $> 1$.}
\label{figmdot}
\end{figure*}

The evolution of the $6 M_{\odot}$ model presented in 
Fig.~\ref{fig6} is characterized by an early phase of strong 
silicates production, followed by a later stage, during which only solid iron forms. 
When silicates production ceases, the mass 
left on the envelope depends on the ratio between the rate of oxygen burning at the 
bottom of the external mantle and the rate of envelope ejection: a smaller 
$\dot M$ renders this behaviour even more extreme, because it increases 
the fraction of silicates--free mass lost by the star. 
This is illustrated in Fig.~\ref{figmdot}. The left panel shows the variations of
the surface content of oxygen and carbon when the \citet{blocker95} (solid lines) and the 
\citet{VW93} (dotted lines) mass loss treatments are used. In both models, 
an initial phase of carbon burning occurs, followed by the depletion of surface
oxygen. The \citet{blocker95} mass loss is so strong that oxygen burning
proceeds at the same rate as surface mantle consumption. Since the \citet{VW93} 
mass loss rate is much slower, the star experiences many more
TPs, and eventually TDU, via an increase in the surface $^{12}$C, favours the
formation of a carbon star. However, carbon dust
production is negligible because surface carbon enrichment is inhibited by HBB.
For the $6 M_{\odot}$ stellar mass model, the mass lost by the star when it is still 
O--rich is $\Delta M \sim 1M_{\odot}$, to be compared to $\Delta M \sim 3M_{\odot}$ in 
the \citet{blocker95} model. As it is shown in the right panel of Fig.~\ref{figmdot},
the production of silicates is consequently smaller and iron is the only dust species produced
for most of the evolution. As a result, the lower mass loss rate leads to a 
reduction of the total dust mass produced by $\sim 2$ dex. 

Dust formation is progressively less affected by the mass loss description as the mass 
of the star decreases, because stars of smaller mass never enter the phase when only 
iron is produced, thus the production of silicates never ceases.

As discussed in Sect. \ref{sec:lowTopacities}, mass loss also plays a role in setting
the stellar mass, which marks the border between silicates producers and carbon--dust 
producers. With our standard choices, based on the \citet{blocker95} treatment, 
we find this threshold mass to be $M \sim 3M_{\odot}$. 
When the \citet{VW93} formulation is adopted, this limit shifts
downwards to $M \sim 2.5M_{\odot}$, because the larger mass left in the envelope for a
given core mass favours HBB ignition.

For SAGBs, \citet{vd11} showed that a mass loss rate increasing with the luminosity, 
as in the \citet{blocker95} recipe, prevents the star from experiencing a very 
advanced nucleosynthesis; on the contrary, in models based on the \citet{VW93} 
formulation only a modest amount of oxygen is left in the envelope for most 
of the SAGB evolution \citep{siess10}. In this latter case the production of 
silicates would be smaller and limited to early evolutionary phases, whereas 
iron would be the main dust species produced. A detailed investigation, aimed 
at understanding the uncertainties affecting stardust from SAGB stars, is in 
preparation and will be presented in a separate paper.

\subsection{The sticking coefficients}
For each dust species, condensation occurs below a threshold temperature
that depends on the differences in the formation enthalpies of the various
molecules entering the reaction associated to the condensation process. Below the
threshold temperature, the destruction rate rapidly declines to zero, thus
leaving the production rate to depend uniquely on the growth--rate coefficients.
The latter, in turn, are linearly proportional to the sticking coefficients, $\alpha$, 
which parametrize the efficiency of the dust formation process.

\begin{figure}
\resizebox{1.\hsize}{!}{\includegraphics{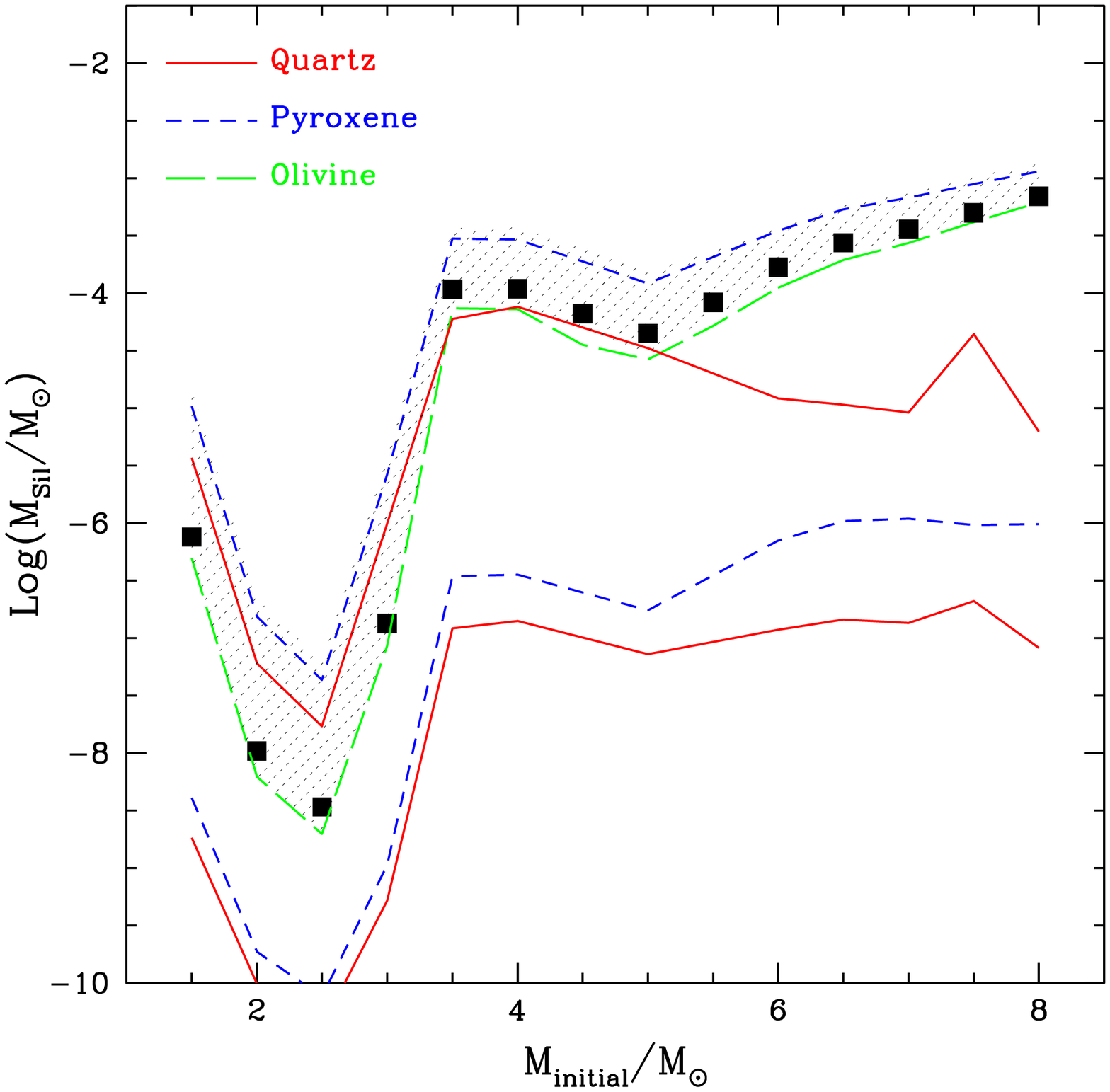}}
\caption{
The filled squares indicate 
the total mass of silicates estimated by the reference model as a function
of the initial stellar mass with the shaded region illustrating the uncertainty
induced by variations of pyroxene and quartz sticking coefficients. 
The dashed (blue) and solid (red) lines indicate, respectively, 
the lowest/highest masses of pyroxene and quartz produced when
the corresponding sticking coefficients are decreased/increased by a factor 5 with
respect to the reference values shown in Table 1. The long--dashed line
indicates the mass of the olivine--type dust produced in the reference model.}
\label{figstick}
\end{figure}

For quartz and pyroxene the sticking coefficients are poorly constrained by laboratory
experiments. Similarly, the exchange coefficient regulating the magnesium fraction 
within olivine and pyroxene is very uncertain. Since for all the stellar models the
surface chemical composition is extremely rich in magnesium, the latter coefficient 
has a negligible effect on dust formation and the equilibria of both hybrid products, 
olivine and pyroxene, are always shifted to the magnesium--rich cases.

To estimate the uncertainties associated to the sticking coefficients we have 
artificially changed by a factor 5 those relative to pyroxene and quartz (see Table 1
for the values adopted in the reference model). The
resulting dust masses are shown in Fig.~\ref{figstick} where dashed (solid) 
lines indicate the mass of pyroxene (quartz) expected when the minimum and maximum
sticking coefficient is adopted. For comparison, we also show the total mass of
silicates expected in the reference model (filled squares) and mass of olivine
(long-dashed line) which is not affected by the uncertainty on the sticking
coefficient and is the most abundant species among silicates. 
Thus, although the pyroxene and quartz masses can vary by a factor $\sim 20$,
their variation has a smaller effect on the total mass of dust: this is 
represented by the shaded region and shows that when the lowest limits for the 
pyroxene and quartz sticking coefficients are used, their contribution is 
completely negligible compared to the olivine, causing a $\sim 30\%$ decrease 
in the total mass of silicates. On the other hand, with the highest sticking
coefficient, pyroxene becomes the dominant species for all stellar models, 
and the silicates mass produced increases by a factor $\sim 3$. 
At present, this quantifies the degree of uncertainty due to
choice of the sticking coefficients on the total dust mass formed.

\section{Discussion}

\begin{figure*}
\begin{minipage}{0.45\textwidth}
\resizebox{1.\hsize}{!}{\includegraphics{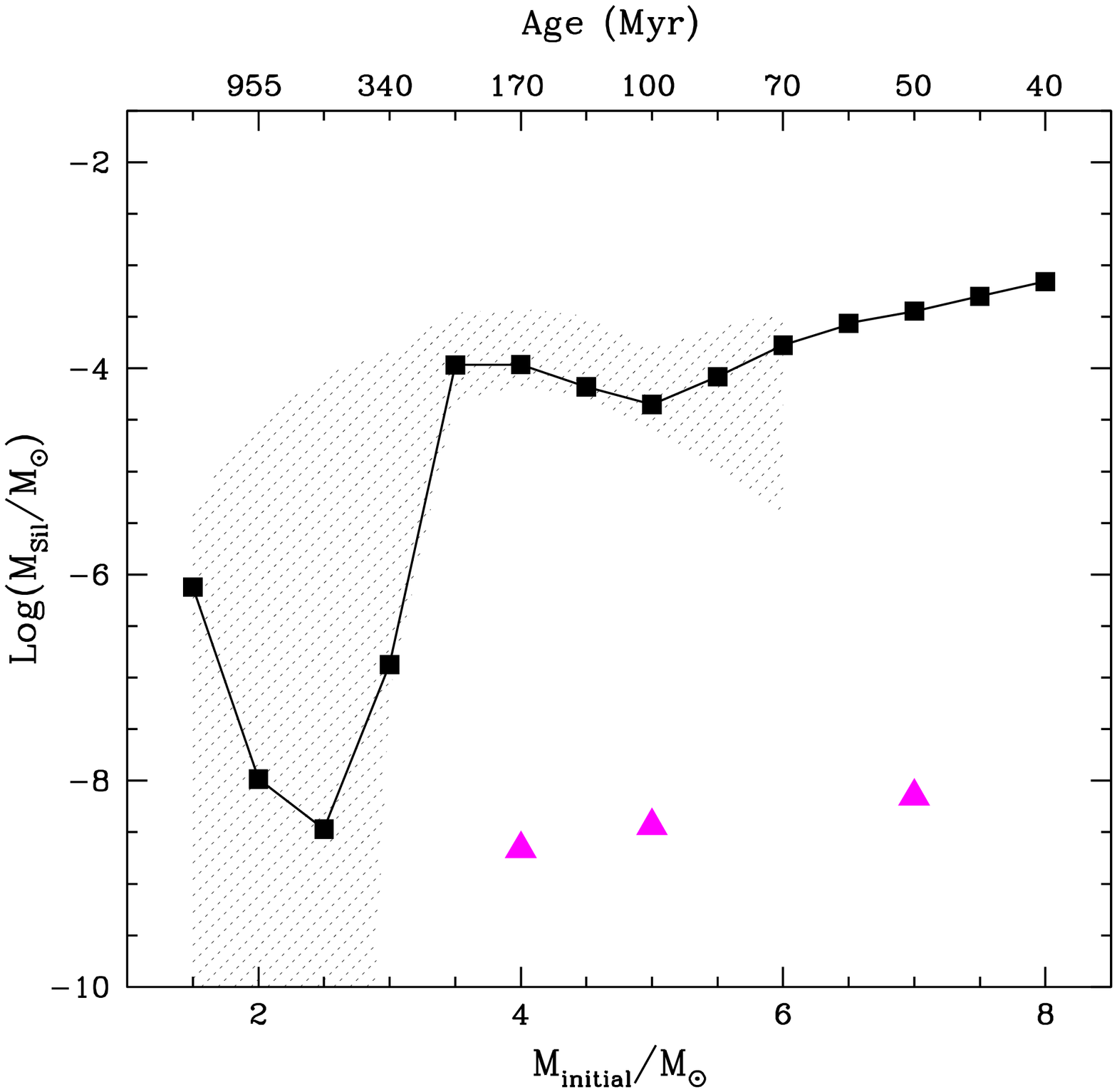}}
\end{minipage}
\begin{minipage}{0.45\textwidth}
\resizebox{1.\hsize}{!}{\includegraphics{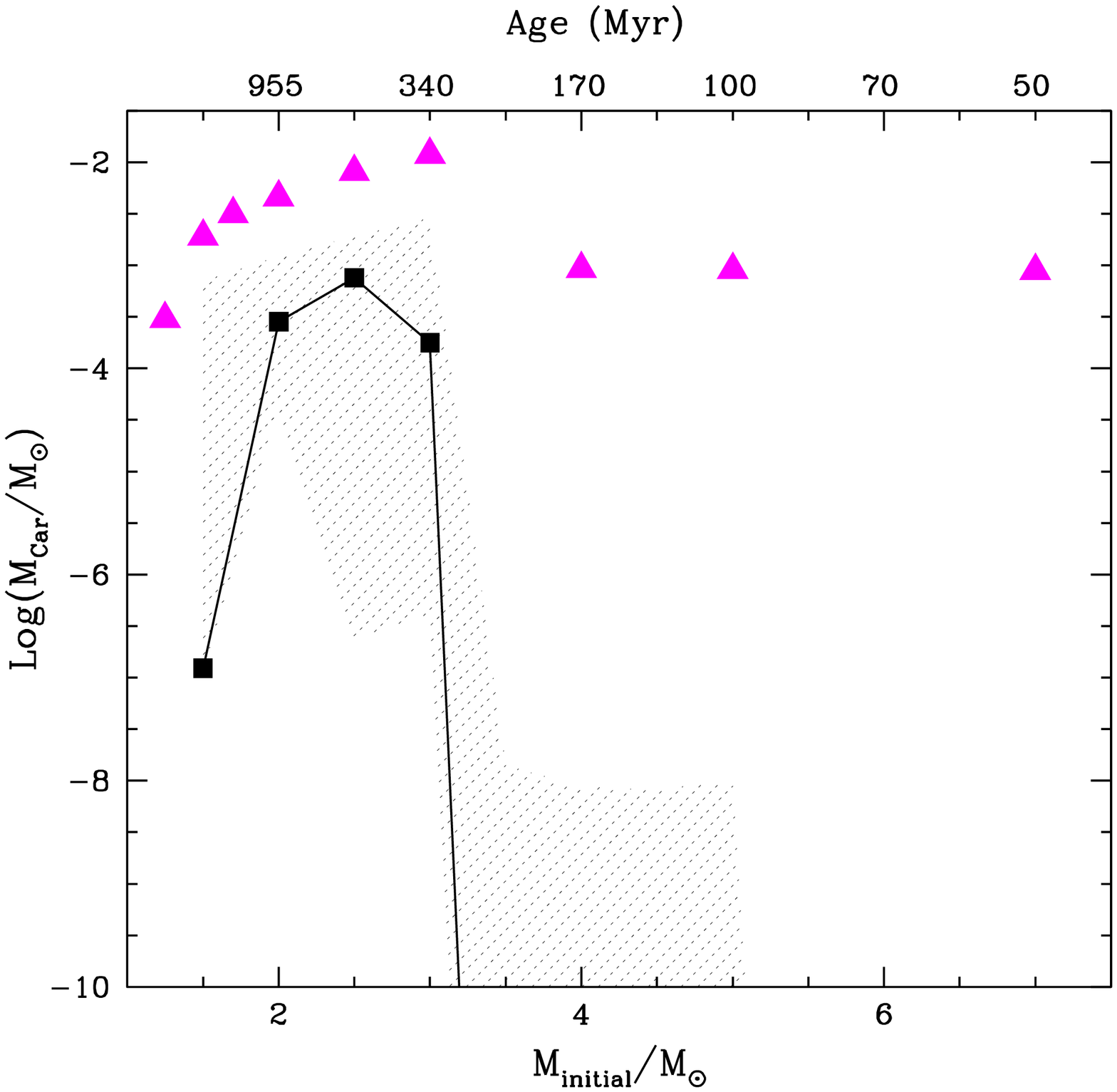}}
\end{minipage}
\caption{The mass of silicates (left) and carbon dust (right) formed around
AGB and SAGB models of a given initial mass. Data points represented by
filled squares indicate the results of our reference AGB and SAGB models. The shaded
areas in both panels illustrate the overall uncertainties associated to these
predictions (see text). For comparison, the dust yields predicted by \citet{fg06} are
also shown (filled triangles). Times on the upper axis indicate for each initial mass 
the duration of the evolutionary phases preceeding the beginning of AGB.
}
\label{figtot}
\end{figure*}

In Fig.~\ref{figtot} we summarize the results of the present study of dust production
by AGBs and SAGBs stars. We show separately the silicates 
(left panel) and carbon (right) dust produced by stars as a function of their
initial mass. The values obtained with the reference model for the AGB evolution 
described in Sect.~\ref{sec:refmodel} are indicated as filled squares. The shaded
areas quantify the uncertainties introduced in dust mass estimates by 
the poorly understood physical processes discussed in
the previous sections. These are limited to stellar masses $M \leq 6M_{\odot}$ because
a detailed investigation of the uncertainties affecting dust production by SAGBs 
will be presented in a future paper. For comparison, the triangles represent
the results obtained by \citet{fg06} using synthetic models of AGBs with initial
metallicity of $Z=0.001$.

The extension of the shaded region in the left panel of Fig.~\ref{figtot} confirms that the
mass of silicates produced by stars with $M \leq 3M_{\odot}$ is extremely uncertain, being
very sensitive to the adopted molecular opacities whose treatment determines whether HBB can
be activated. In our standard case, models within this range of masses are precited to
produce carbon--type dust, with a poor production of silicates: this explains the almost
null lower limit in the left panel of Fig.~\ref{figtot}.
On the other hand, use of classic opacities, coupled with the \citet{VW93} mass loss prescription, 
allows HBB conditions to be reached even for $M \sim 2 M_{\odot}$, thus preventing the envelope of 
the star to become carbon--rich, and favouring the production of silicate--type dust. 
These assumptions, coupled with the highest choice of the sticking coefficients for the 
formation of pyroxene and quartz grains (see Fig.~\ref{figstick}), determine the
upper envelope of the shaded area in the low--mass regime of Fig.~\ref{figtot} (left panel). 

For stars experiencing HBB, the uncertainties on silicates production are smaller.
The lower limit on the shaded regions is set by the mass loss modelling, whereas the upper 
limit is due to the choice of the sticking coefficients. In agreement with the discussion of the 
previous sections, we find that the most uncertain models are those experiencing
strong HBB ($M \sim 6 M_{\odot}$).

The right panel of Fig.~\ref{figtot} shows that carbon dust production is also uncertain
in the low--mass regime (M$\leq 3M_{\odot}$), due to the combined effects of mass loss and opacities for 
carbon--rich mixtures, that are essential in determining whether HBB
conditions are reached. In our reference model no carbon dust is expected for $M > 3.5M_{\odot}$,
whereas in models calculated with the \citet{VW93} mass loss treatment some production is
expected, because the envelope is consumed so slowly that eventually TDU can lead to the
C/O$>1$ condition. This determines the extent of the shaded region associated to stars of $4 M_{\odot}$ 
in the right panel of Fig.~\ref{figtot}.

A comparison between the data marked with squares and triangles in Fig.~\ref{figtot} shows that 
the dust yields based on synthetic AGB models \citep{fg06} can not be reconciled with our results, 
even taking into account the uncertainties on the physical processes discussed so far. 
According to \citet{fg06}, the dominant dust species produced by stars with $Z=0.001$ and initial
stellar masses $1 M_{\odot} \le M \le 7 M_{\odot}$ is carbon--dust; the production of silicates 
is limited to stars with initial masses $\ge 4 M_{\odot}$ but amounts to a few percent of the
total dust mass formed.

These differences can be entirely ascribed to the adopted description of the convective
transport in the innermost regions of the convective envelope. As shown by \citet{vd05a},
use of the FST model leads to stronger HBB, at odds with the MLT treatment, that favours 
a modest HBB only in the most massive stars. In the latter case, independently of the mass,
TDU eventually dominates and the carbon--star stage is reached. It is not surprising that
the results by \citet{fg06} predict the formation of C--type dust; in our case this is possible
only for models not experiencing HBB, i.e. for M$\leq 3M_{\odot}$.

In the low--mass regime the results are in qualitative agreement, the carbon--dust mass presented
by \citet{fg06} being larger; this is partly due to the different choices for the low--T molecular
opacities and the mass loss description, and also the efficiency of the TDU extension, which
in our case is found by neglecting any overshoot from the base of the envelope, and is thus to
be considered as a conservative estimate.

\subsection{Implications for cosmic dust enrichment}

It is interesting to investigate the implications of the dust yields
for AGB and SAGB stars within the context of cosmic dust enrichment.

Following the analysis done by Valiante et al. (2009), we can estimate
the time evolution of the dust mass (and of individual dust species) 
produced by a stellar population as,
\begin{equation}
M_{\rm dust}(t) = \int_0^{t} dt' \int_{m_\ast(t')}^{m_{\rm up}} m_{\rm dust}(m) \phi(m) SFR(t'-\tau_m) dm,
\label{eq:dustyield}
\end{equation}
\noindent
where $m_{\rm dust}(m)$ are the stellar mass dependent dust yields, $\phi(m)$ is the stellar initial
mass function (IMF), $m_{\rm up}$ is the upper limit of the stellar mass range, $\tau_m$ is the
lifetime of a star with mass $m$, $SFR$ is the star formation rate and $m_\ast(t')$ is the mass of
a star with lifetime $\tau_{m_\ast} = t'$. 
We adopt a Larson IMF (Larson 1998) which follows a Salpeter--like power--law at the high-mass end
but flattens below a characteristic stellar mass, $m_{\rm ch} = 0.35 M_\odot$,
\begin{equation}
\phi(m) \propto m^{-(\alpha+1)} e^{-m_{\rm ch}/m},
\end{equation}
\noindent
where $\alpha=1.35$ and the IMF is normalized between $m_{\rm inf} = 0.1 M_{\odot}$ and $m_{\rm up} = 100 M_{\odot}$.
For stars with masses in the range $1 M_{\odot} \le m \le 8 M_{\odot}$, we take the AGBs and SAGBs dust yields predicted
by the reference model discussed in Sec.~\ref{sec:results}. At larger stellar masses, dust can be synthesized in the 
ejecta of core--collapse SNe and dust yields as a function of the progenitor mass and metallicities have been
published in the literature (Kozasa, Hasegawa \& Nomoto 1991; Todini \& Ferrara 2001; Nozawa et al. 2003; Bianchi \& Schneider
2007). Here we neglect this contribution to dust production and we only consider 
the AGBs and SAGBs dust yields. 

The time evolution of the total dust mass, described by eq.~\ref{eq:dustyield}, depends also on the 
stellar lifetimes and on the adopted star formation history. For low and intermediate mass stars, lifetimes
are computed directly from the ATON evolutionary models and range between $72.4$~Myr and $1.82$~Gyr for
AGBs and between 41.7 and 61.7 Myr for SAGBs. 

\begin{figure}
\resizebox{1.\hsize}{!}{\includegraphics{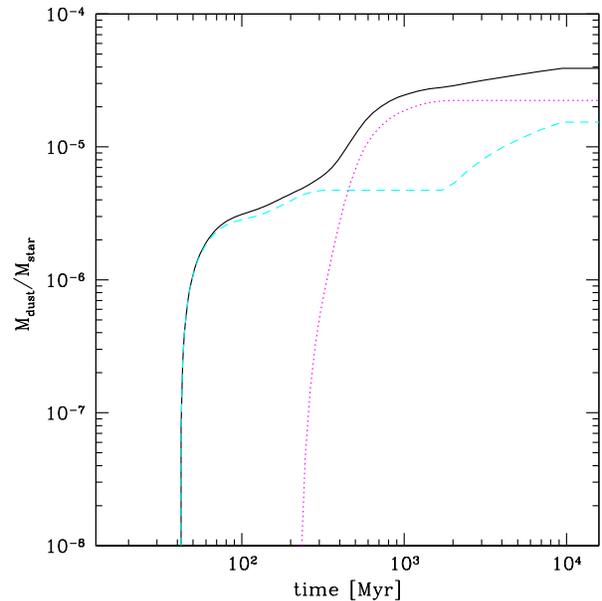}}
\caption{Time evolution of the mass of dust produced by AGBs/SAGBs normalized to the total mass of stars (solid line). 
All stars are assumed to form in a single burst at $time = 0$ with a metallicity of $Z= 5\times 10^{-2} Z_{\odot}$ and a 
Larson IMF with $m_{\rm ch} = 0.35 M_{\odot}$. The dashed and dotted lines show the separate contributions of silicates 
and of carbon dust, respectively.}
\label{fig:dustyield}
\end{figure}

Fig.~\ref{fig:dustyield} shows the evolution of the total mass of dust (solid line), silicates (dashed line)
and carbon grains (dotted line). All the stars are assumed to form in a single burst at $t=0$ with initial 
metallicity of $Z=5\times 10^{-2} Z_{\odot}$ and the mass of dust is normalized to the total stellar mass formed. 

%
Since massive AGBs and SAGBs stars are the most efficient silicates producers, about $50$~Myr after the burst
silicates dominate the dust mass evolution. When stars of $M \sim 3 M_{\odot}$ have ended their evolution,
about 300 Myr after the burst, the production of silicates ceases and restarts only when stars with $M \le 2 M_{\odot}$ 
reach their AGB phase ($\sim 1.7$~Gyr after the burst). Carbon dust enrichment is delayed to longer timescales, 
being produced only by stars with masses $\le 3 M_{\odot}$; this represents the dominant dust component
after $\sim 470$~Myr of the burst.
 
Thus, if AGBs and SAGBs stars are the dominant stardust sources, the ISM in galaxies at $z > 10$ and 
in young starbursts with ages $< [450 - 500]$~Myr at any redshift will be predominantly enriched by 
silicates and no carbon dust features are expected to be observable. 
It is clear that these findings may partially depend on the adopted initial metallicity of the stars,
which affect both the core mass and the surface chemistry of the stars.
In a future study, we will investigate the formation of dust using a grid of AGBs/SAGBs exploring the
full range of metallcities $0 \le Z \le Z_{\odot}$. 

Presence of carbon dip in the (rest--frame) UV spectrum of galaxies at z$\sim 1-3$ \citep{noll07, noll09}
suggests that these host evolved (age exceeding a few hundred million years) stellar populations.
The lack of carbon dip in the spectra of AGN and high redshift quasars, accompanied by the presence
of the silicate feature at 10$\mu$m, suggests that these active nuclei are hosted in young stellar
populations \citep{maiolino01, lutz08, maiolino04, gallerani10}.


\section{Conclusions}
We investigate the production of dust using physical models of AGB and SAGB stars with mass in the range
$1 M_{\odot} \leq M \leq 8M_{\odot}$ and initial metallicities $Z=0.001$. 

The type of dust formed depends on the HBB ignition at the bottom of the external mantle: 
HBB is accompanied by the depletion of surface carbon, which, in turn, 
prevents the production of carbon dust, in favour of silicates. 
The threshold mass above which only silicates are formed is found to be $M = 3M_{\odot}$.
Below this limit, the evolution in the surface chemistry is dominated by TDU with the
consequent formation of carbon dust.

The distribution with stellar mass of the amount of dust formed shows a minimum corresponding
to the $[3.5 - 5] M_{\odot}$ mass range due to the combined effects of the soft HBB experienced
by these stars and inefficient mass loss. The strength of HBB increases with stellar mass leading
to almost complete destruction of surface oxygen for stars with $M \sim 6 M_{\odot}$; this, in turns,
limits the formation of silicates and leaves solid iron as the dominant dust species. 

SAGB models are expected to produce great amounts of silicates. They experience large 
mass loss rates, ejecting their external envelope before significant surface oxygen depletion.

In contrast with previous studies, we show that use of the FST treatment, as in the present 
investigation, leads to an efficient HBB ignition in all models with masses $M>3M_{\odot}$, that 
are thus predicted to produce silicates.This is at odds with previous results published in the 
literature coming from synthetic AGB models based on the MLT treatment, which predicts C--rich dust 
to be the dominant species at all masses.

As a consequence of the strongly mass--dependent dust composition found by our physical AGB models
based on FST treatment of turbulent convection, whenever the ISM dust enrichment is dominated by
AGBs and SAGBs at low metallicities, we predict an early phase ($\sim 300$~Myr) of silicate production
followed by the formation of C--rich dust.

%

\section*{Acknowledgments}
The authors are indebted to Prof.~Gail for the many helpful discussions, and for
providing the data for the formation enthalpies of some of the molecules entering this
treatment. MDC acknowledges finantial support from the Observatory of Rome.

\end{document}